%% file: main.tex
\documentclass[preprint,12pt]{elsarticle}
\usepackage{graphicx} 
\graphicspath{ {./figures/} }
\usepackage{float}
\usepackage{multirow}
\usepackage{hyperref}
\usepackage{array}
\usepackage[many]{tcolorbox} 
\definecolor{main}{HTML}{5989cf}
\definecolor{sub}{HTML}{cde4ff}

\usepackage{longtable}
\usepackage{array}
\usepackage{booktabs} 
\usepackage{multirow}
\usepackage{ragged2e}
\usepackage{makecell}
\usepackage{lscape}

\title{Use of Peer Code Review to Identify Potential Security Vulnerabilities in Open Source Software: A Systematic Literature Review}

\author[1]{Md Niaz Morshed\corref{cor1}}
\ead{mmorshed@crimson.ua.edu}
\author[1]{Jeffrey C. Carver}
\ead{carver@cs.ua.edu}

\affiliation[1]{organization={Department of Computer Science},
addressline={The University of Alabama},
postcode={35487},
city={Tuscaloosa},
country={USA}}

\date{October 2025}

\begin{document}

\begin{abstract}
\input{sec-Abstract}
\end{abstract}

\maketitle
\begin{sloppypar}

\section{Introduction}
\label{sec:Introduction}
\input{sec-Intro}

\section{Background}
\label{sec:Background}
\input{sec-Background}

\section{Methodology}
\label{sec:Methodology}
\input{sec-Methodology}

\section{Results}
\label{sec:Result}
\input{sec-Result}

\section{Discussion}
\label{sec:Discussion}
\input{sec-Discussion}

\section{Threats to validity}
\label{sec: Threats to validity}
\input{sec-Threat_to_validity_new}

\section{Conclusion}
\label{sec:Conclusion}
\input{sec-Conclusion}

\bibliographystyle{elsarticle-num} 
 \bibliography{cas-refs}

 \section{Appendix}
\label{sec:Appendix}
\input{sec-Appendix}

\end{sloppypar}
\end{document}

%% file: sec-Abstract.tex
\textbf{Context:} 
Security is a major software development concern.
Security vulnerabilities in Open-Source Software (OSS) can have severe consequences. 
OSS project participants widely employ peer code review to increase code quality and prevent security flaws.
However, even with mandatory code reviews, many OSS projects still contain security flaws.
The presence of these flaws highlights the importance of evaluating the effectiveness of code reviews in addressing security issues.

\vspace{8pt}
\noindent
\textbf{Objectives:} 
This Systematic Literature Review analyzes how peer code review helps identify OSS security vulnerabilities. 
The study synthesizes existing evidence on vulnerability types identified or missed, current security practices, tool supports, reviewer challenges, and factors enhancing detection.

\vspace{8pt}
\noindent
\textbf{Method:}
We conducted a search across five major digital libraries. 
We screened 1067 studies and selected 20 primary studies using predefined inclusion and exclusion criteria. We analyzed the data using qualitative analysis and categorized vulnerabilities using the Common Weakness Enumeration (CWE) framework.

\vspace{8pt}
\noindent
\textbf{Results:}
Peer code reviews effectively detect low-hanging, localized, well-known vulnerabilities (e.g., buffer overruns, race conditions, crashes, XSS) but tend to miss subtle, complex, context-dependent issues (e.g., input validation weaknesses, access control weaknesses, resource lifecycle weaknesses).
Effective security practices are tied to reviewer expertise, familiarity with the codebase, frequency of small code changes, and organizational support.
Security analysis tools, particularly Static Application Security Testing (SAST) tools, reduce reviewer workload and improve detection, but cannot replace human judgment.
Reviewers still suffer from constraints of security knowledge and experience, time, and comprehending complex systems.
Reviewer effectiveness improves with security training, domain experience, contextual familiarity, formalized review processes, and favorable review environments.

\vspace{8pt}
\noindent
\textbf{Conclusion:}
Peer code review contributes to OSS security, but is not sufficient alone. 
Strengthening the effectiveness of peer code review requires integrating structured processes and checklists, security training, organizational incentives, and lightweight automated tool support to help reviewers detect subtle, complex vulnerabilities, and reinforcing secure development practices.

%% file: sec-Intro.tex
Security vulnerabilities are a significant concern for software systems. 
Their exploitation can result in significant financial losses, privacy violations, and service disruptions. 
The impact of these vulnerabilities becomes even more pronounced if discovered after software deployment. 
Notable examples, including the Equifax~\cite{Equifax} breach, which exposed sensitive personal information, and the Heartbleed~\cite{Heartbleed_2014} vulnerability, which enabled attackers to read sensitive memory, highlight the real-world impact of security vulnerabilities in widely used Open Source Software (OSS). 
The true prevalence of vulnerabilities is likely higher, as many remain undetected.
In 2024, NIST reported over 39,000 vulnerabilities~\cite{NIST_report}, underscoring the need for early detection and mitigation.

OSS has become a cornerstone of modern software development, fostering innovation and promoting collaboration across global communities. 
OSS development relies heavily on distributed, volunteer-driven collaboration, in which contributors have varying levels of expertise. 
This diversity of contributor skill level can lead to inconsistent code quality and increased vulnerabilities. 
Unlike proprietary projects with dedicated security teams, OSS relies on community members to identify and address security vulnerabilities. 
These vulnerabilities can arise during development and remain undetected, posing significant risks to users and organizations. 
Due to the prevalence of OSS, these vulnerabilities can have wide-ranging effects.

Developers commonly use automated program analysis tools to support vulnerability detection~\cite{ashisandjagannath2013}.
These tools employ static techniques to identify insecure code patterns and dynamic techniques to expose runtime failures. 
However, prior studies have shown that such tools suffer from limitations, including high false-positive rates in static analysis and scalability challenges in dynamic analysis~\cite{johnsonandrobert2013,arthoandarmin2005}.
These limitations restrict their effectiveness when used in isolation.

Peer code review (a.k.a code review) is a widely adopted collaborative practice in which developers manually inspect code changes to improve quality and detect defects, including security vulnerabilities~\cite{Ackerman1984,ackerman1989}.
Unlike automated tools, peer review incorporates human judgment and contextual understanding, which are essential for identifying subtle or context-dependent security issues.
Previous research indicates that code reviews are crucial for identifying and resolving security issues early on and contribute positively to the development of secure software~\cite{mcgraw2006,meneelyatel2014,shinetal2011,thompsonetal2017}.
An effective code review can reveal approximately 50\% of the security issues present~\cite{chessandjacob2007}.
Consequently, many OSS and industrial projects mandate code review as part of their development workflow~\cite{peterandbird2013}.
Despite this widespread adoption, security vulnerabilities continue to escape detection during mandatory reviews and surface after release~\cite{rajpauletal2021}.

Existing research on code review has largely focused on general software quality, bug detection, and adherence to coding standards~\cite{Aurumandwohlin2002,mcintosh2014,bosuandcarver2017}.
Security-specific outcomes have received comparatively less attention, even though security vulnerabilities differ from general defects in their causes, impact, and detection requirements~\cite{camiloandmeneely2015,shahedandhassan2011,meneelyandsrinivasan2013}. 
Although several empirical studies have demonstrated the potential of code reviews to improve security~\cite{carverandbosu2014,edmundsonetal2013,Munaiahandmeyers2017}, it remains unclear how effective peer code reviews are in identifying security issues in OSS.

To address this gap, this paper presents a systematic literature review of code review for identifying potential security vulnerabilities in open-source software. 
The review synthesizes empirical evidence on the types of potential vulnerabilities detected and missed during reviews, existing security-oriented review practices, the role of automated security tools, challenges faced by reviewers, and factors that enhance the detection of potential vulnerabilities.
By consolidating findings across studies, this work aims to clarify the current state of knowledge and identify limitations that hinder effective security-focused code reviews in OSS.

The key contributions of this study are as follows:
\begin{itemize}
 \item A structured categorization of potential security vulnerability types that are commonly identified or missed during code reviews in OSS.
 \item A synthesis of current security practices applied during code reviews in OSS.
 \item An assessment of how security analysis tools support potential security vulnerability detection during code review.
 \item An identification of key challenges and limitations that hinder reviewers from effectively detecting potential security vulnerabilities.
 \item An analysis of factors that enhance reviewers’ ability to detect potential security vulnerabilities.
 \end{itemize}
 
The research artifacts associated with this systematic literature review, including the data extraction forms and the raw data of identified and missed potential vulnerabilities are provided as supplementary material via an online repository\footnote{\url{https://zenodo.org/records/18945199}} to ensure transparency and reproducibility.

The paper is organized as follows: Section \ref{sec:Background} provides background information. 
Section \ref{sec:Methodology} outlines the research methodology used for our systematic literature review.
Section \ref{sec:Result} analyzes the results obtained. Discussion derived from our study is presented in Section \ref{sec:Discussion}. 
Threats to validity are examined in Section \ref{sec: Threats to validity}.
Finally, Section \ref{sec:Conclusion} concludes the paper.

%% file: sec-Background.tex
This section provides the necessary background for the study by examining key topics in software security and peer code-review practices in OSS.

\subsection{Software Vulnerabilities}
\label{sec_security_vul}

A software vulnerability is a flaw or weakness in software that attackers can exploit to compromise the confidentiality, integrity, or availability of a system~\cite{OWASPAbout2013}.
Such vulnerabilities can lead to serious consequences, including data breaches, financial loss, and reputational damage~\cite{Telangetal2007}. 
To reduce these risks, organizations increasingly adopt a “shift-left” approach, which emphasizes detecting security issues earlier in the development lifecycle, including during code review~\cite{boehmetal2001}.

Numerous security vulnerabilities can arise at various stages, including implementation, design, and system architecture. 
To support consistent analysis and communication, the security community categorizes vulnerabilities using the MITRE Common Weakness Enumeration (CWE) framework~\cite{mitre_cwe}. 
CWE provides a standardized taxonomy in which each weakness is assigned a unique identifier and description, enabling systematic classification and comparison across studies.
Software developers and researchers discuss security issues at different levels of abstraction. 
A \textit{weakness} refers to a condition in software that may contribute to vulnerabilities under certain circumstances, whereas a \textit{vulnerability} refers to a concrete and exploitable security flaw in a deployed system. 
Security databases such as Common Vulnerabilities and Exposures (CVE) typically catalog vulnerabilities after deployment, whereas developers and reviewers can identify weaknesses earlier during development activities such as peer code review. 
This distinction is particularly important in peer code review because reviewers examine insecure coding patterns and potential Vulnerabilities before deployment.
In this review, we adopt the CWE framework to categorize the potential security vulnerabilities reported in the primary studies.

\subsection{Peer Code Review}

Peer code review is a collaborative practice in which developers other than the author examine code changes before they are integrated into the codebase.
This process is typically iterative and continues until reviewers approve the code change for integration.
Reviewers aim to detect defects, enforce coding standards, and improve software quality~\cite{bacchbird2013}. 
The practice originated from a systematic code review method known as code inspections~\cite{fagan1976}. 
This approach followed a structured and synchronous process where trained teams reviewed code through formal meetings.

Modern code review (MCR), introduced by Cohen~\cite{cohen2010modern} and later studied by Bacchelli and Bird~\cite{bacchbird2013}, is a widely adopted form of code review.
MCR is typically informal, tool-based, asynchronous, and focused on reviewing incremental code changes rather than entire codebases~\cite{bacchbird2013}.
Tool support allows reviewers to inspect code diffs, leave comments, and iterate on changes efficiently.

\subsection{Peer Code Review Practices and Security}

Beyond general quality assurance, peer code review also plays an important role in software security. 
Prior studies show that code reviews can help identify certain classes of vulnerabilities and support knowledge sharing among developers~\cite{mantylaandcasper2009,bellerandbacchelli2014,bacchbird2013}. 
In addition, McGraw emphasized that peer code review is crucial for identifying and fixing security vulnerabilities~\cite{mcgraw2006}.
Howard~\cite{howard2006} further highlighted that effective security-focused code review requires reviewers to understand common vulnerability patterns and prioritize high-risk code areas, such as legacy code, externally accessible components, code with a history of vulnerabilities, complex code, code that runs by default, and modules handling sensitive data.
Many large OSS projects, including Chromium, Android, Qt, and Mozilla, as well as major industrial organizations, have adopted mandatory peer code review as part of their development processes~\cite{peterandbird2013,rajpauletal2021}.

Peer code review has received considerable attention for its role in improving software quality~\cite{bacchbird2013, mcintosh2014, Thongtanunametal2015}.
Edmundson et al.~\cite{edmundsonetal2013} demonstrated that although peer code review enhances software security, developers are not always able to identify all potential vulnerabilities.  
Although this practice improves security, some vulnerabilities remain undetected.
This gap motivates a systematic examination of how peer code review contributes to security in OSS. 
Therefore, this research aims to conduct a Systematic Literature Review (SLR) examining peer code reviews to detect potential security vulnerabilities in open-source software.

%% file: sec-Methodology.tex
We followed the guidelines of Kitchenham and Charters~\cite{kitchenham2007slrguidelines} to conduct our SLR.
The following subsections detail the process we followed.

\subsection{\textit{Defining the research questions}}

We want to dive deep into the latest research on how peer code reviews can help uncover security vulnerabilities in OSS.
Conducting an in-depth analysis of peer code review research to identify security vulnerabilities within OSS is valuable for researchers seeking to understand existing exploration and identify unresolved questions.

Investigating the types of security vulnerabilities that can be identified during code review in OSS is essential to mitigate potential security risks at an early stage of development~\cite{McGraw2004}. 
Delays in detecting and resolving security vulnerabilities increase their associated costs over time~\cite{MacGrwa_2008}. 
Hence, early identification and resolution of vulnerabilities through meticulous code reviews can significantly reduce the overall cost of developing secure software. Therefore, the first research question is:

\begin{tcolorbox}[enhanced,drop shadow]
\textbf{RQ1 What types of potential security vulnerabilities can be identified or missed during peer code review in OSS?}
\end{tcolorbox}

OSS projects typically involve a diverse group of contributors, which can lead to inconsistent code quality and the introduction of potential security vulnerabilities. 
The role of security practices during code reviews in OSS is crucial for detecting and mitigating these vulnerabilities~\cite{carverandbosu2014, mcgraw2006}. 
Although several guidelines and frameworks for secure coding practices have been suggested~\cite{mcgraw2006, OWASPAbout2013}, there is a significant gap in empirical research that investigates the actual code review practices used in real-world scenarios. 
To provide a broad overview of security code review practices, the second research question is as follows. 

\begin{tcolorbox}[enhanced,drop shadow]
\textbf{RQ2	What are the current security practices during peer code review in OSS?}
\end{tcolorbox}

Security analysis tools, including static and dynamic analysis tools and vulnerability scanners, play a crucial role in the automated detection of security weaknesses, vulnerabilities, and code patterns. 
Previous studies have shown that incorporating these security analysis tools into the code review process can substantially improve the detection of potential vulnerabilities~\cite{charoenwet2024, alfadelandandnagy2023}.
To explore the impact of security analysis tools on the code review process in OSS, the third research question is formulated as follows.

\begin{tcolorbox}[enhanced,drop shadow]
\textbf{RQ3 How do security analysis tools contribute to improving OSS security during peer code reviews?}
\end{tcolorbox}

Investigating the challenges reviewers face in ensuring security during code review is crucial due to the unique dynamics of OSS development. 
Unlike proprietary software, OSS relies on voluntary contributions and peer review. 
The literature highlights that reviewers often struggle to identify complex vulnerabilities due to a lack of expertise, security knowledge gaps, cognitive overload, and time constraints ~\cite{carverandbosu2014,brazandbacchelli2022,meneelyatel2014}.
Understanding these challenges is essential for developing effective methodologies, tools, and practices to enhance security-focused code reviews in OSS.
So the fourth research question is as follows.

\begin{tcolorbox}[enhanced,drop shadow]
\textbf{RQ4 What challenges and limitations do reviewers face in ensuring security during peer code reviews in OSS?}
\end{tcolorbox}

Identifying security vulnerabilities during code reviews is crucial for maintaining OSS security. 
The existing literature identifies several factors that enhance a reviewer's ability to identify vulnerabilities, including domain knowledge, security awareness, experience with secure coding practices, security knowledge, familiarity with the code base, review time, and the integration of automated tools~\cite{rajpauletal2021,dibiaseatal2016,brazandchristian2022,carverandbosu2014,meneelyatel2014}.
Despite existing insights, a deeper understanding of these enhancing factors is essential to improve the effectiveness of security-focused code reviews. 
Therefore, the fifth research question is as follows.

\begin{tcolorbox}[enhanced,drop shadow]
\textbf{RQ5 What factors enhance a reviewer’s ability to detect potential security vulnerabilities during peer code reviews in OSS?}
\end{tcolorbox}

\subsection{\textit{Data source selection}} 
We searched the following databases commonly used in SE SLRs\cite{Kuhrmann2016,kitchenham2007slrguidelines}.

\begin{itemize}
  \item IEEE Xplore
  \item ACM Digital Library
  \item ScienceDirect
  \item Springer Link
  \item Scopus
\end{itemize}

\subsection{\textit{Search string}} 
To find relevant studies, we selected three terms to serve as the basis for the search. 
First, we chose "peer code review." 
Second, we chose "security" because it is the focus of our study. 
Third, we chose "open-source software" to exclude work focused on industrial software, which might employ different security practices.
We then form the search strings by combining the keywords and their synonyms.
We identified synonyms for each main term by reviewing prior related literature reviews and primary studies in software engineering and security. 
We refined and finalized the search terms and search string through discussion among the authors.
This process ensured that our search terms captured the commonly used variations of the core concepts.

\begin{itemize}
  \item \textbf{Term 1:-}  "peer code review" OR "code review" OR "code inspection" OR "modern code review" OR "code checking" OR “code analysis”
  \item \textbf{Term 2:-} "security" OR "vulnerability" OR "threat" OR "attack" OR "vulnerable" OR “secure”
  \item \textbf{Term 3:-} "open-source software" OR "OSS" OR "FOSS" OR "free open-source software" OR "non-proprietary software" OR “FLOSS”
\end{itemize}

Because all three terms are necessary to identify the correct papers, we formed the search string as follows:  \textbf{(Term 1) AND (Term 2) AND (Term 3)}.

\begin{tcolorbox}[enhanced,drop shadow]
\textbf{Search String: ("peer code review" OR "code review" OR "code inspection" OR "modern code review" OR "code checking" OR “code analysis” ) AND ("security" OR "vulnerability" OR "threat" OR "attack" OR "vulnerable" OR “secure” ) AND ("open-source software" OR "OSS" OR "FOSS" OR "free open-source software" OR "non-proprietary software" OR “FLOSS” )}
\end{tcolorbox}

\subsection{\textit{Selection criteria}} 

We applied the inclusion and exclusion criteria listed in Table~\ref{table: Inclusion and exclusion} during screening.
For a study to be included, it must meet at least one inclusion criterion and not meet any exclusion criteria. 

\begin{table}[!htb]
\centering
\input{tab_Inclusion-Exclusion}
\caption{Inclusion and exclusion criteria}
\label{table: Inclusion and exclusion}
\end{table}

\subsection{\textit{Primary study selection}}
We tailored our search string to match the specific syntax requirements of each digital library and executed the search in May 2023.
This search retrieved 941 records across the five databases (Table~\ref{table:Search_DB}).
Of those 941 papers, 9 appeared in more than one database.
Once we removed the duplicates, 932 unique records remained for screening.

To maintain currency, we performed a follow-up search in September 2024 using the same search strategy. 
The updated search retrieved 126 additional records published between May 2023 and September 2024 (Table~\ref{table:Search_DB}).
Similar to the initial search, of the 126 new papers, 2 appeared in more than one database. 
After removing duplicates, 124 unique records remained for screening.
We processed the initial and updated search rounds independently during duplicate removal and screening, and then combined the final screening results from both rounds.

\begin{table}[!htb]
\centering
\input{tab_Search_DB}
\caption{Search Results by Database}
\label{table:Search_DB}
\end{table}

Study selection proceeded in four phases: (1) removal of duplicates, (2) title and abstract screening, (3) introduction and conclusion screening, and (4) full-text review. 
Table~\ref{table: Inclusion and exclusion} summarizes the inclusion and exclusion criteria. 
To ensure reliability, the second author independently applied the inclusion and exclusion criteria to a 20\% random sample of papers in each phase. 
Inter-rater disagreements occurred in 15\%, 8\%, and 5\% of papers during the second to fourth screening phases, respectively.
All disagreements were resolved through discussion until the authors reached full agreement. 
Finally, we performed backward snowballing on the reference lists of the included studies. This process did not identify any additional papers meeting our criteria.

\begin{table}[!htb]
\centering
\input{tab_Study_Selection}
\caption{Study Selection Process}
\label{table:Study_Selection}
\end{table}

The 20 primary studies span 2009–2024, but most evidence is recent. 
Nineteen studies (95\%) were published in 2013 or later, and among them 11 were published between 2020 and 2024. 
Table \ref{tab:primary_studies} in the Appendix reports publication year and author affiliations for each study. 
The affiliations cover multiple research groups across regions (e.g., Australia, Canada, Europe, and the United States), and we did not observe a single dominant group.

Most included studies are empirical (19/20, 95\%).
The remaining study is a systematic review that synthesizes empirical evidence related to modern code review practices.
These studies mainly analyzed OSS artifacts such as review comments, pull requests, patches, and review discussions mined from real-world open source projects and ecosystems (e.g., GitHub, Gerrit/Chromium OS, Qt, OpenStack, Mozilla Firefox, SmartSHARK, npm, and Python packages).
Thus, our synthesis is grounded in evidence from real-world OSS development data, even though we do not directly mine repositories in this review.

\subsection{\textit{Data extraction}}
\label{Data_extraction}
The objective of this stage is to design data extraction forms. 
The first author then extracted all the data for this publication using that form.
Then, the second author evaluated the extracted data to check for accuracy.

\begin{table}[!htb]
\centering
\input{tab_Data_Extraction}
\caption{Data Extraction Form}
\label{table:Data_Extraction}
\end{table}

\subsection{\textit{Data Synthesis}}

For each primary study, we extracted findings relevant to each research question using the structured data extraction form described in Section~\ref{Data_extraction}.
The extracted data included reported vulnerability types, security practices, tool usage, reviewer challenges, and enhancing factors, depending on which research questions the study addressed. 
This process produced a structured dataset of findings organized by research questions across all included primary studies.

We grouped extracted findings into recurring themes by identifying conceptually similar findings across studies using thematic aggregation.
For RQ1, we mapped the reported vulnerability types and weaknesses to higher-level categories using the MITRE CWE-1000 Research Concepts view through a bottom-up aggregation process, as described in Section \ref{security_Classification}. 
This process enabled consistent aggregation and comparison of security issues reported across different studies.
For RQ2 through RQ5, we iteratively compared and grouped conceptually similar findings across studies to derive higher-level themes and categories.
Within each theme, we synthesized findings across studies by identifying recurring patterns, similarities, differences, and complementary observations reported in the literature. 
Both authors reviewed the synthesized themes and resolved disagreements through discussion to improve consistency and reduce interpretation bias.

\vspace{8pt}

%% file: tab_Inclusion-Exclusion.tex
\begin{tabular}{|p{7cm}|p{7cm}|}
\hline
\textbf{Inclusion Criteria} & \textbf{Exclusion Criteria} \\ \hline

Articles that examine security vulnerabilities during peer code review in OSS 
& Duplicate articles \\ \hline

Articles that synthesize security practices applied during peer code review in OSS 
& Articles that are not peer-reviewed \\ \hline

Articles that discuss security analysis tools supporting vulnerability detection during code reviews in OSS 
& Articles not written in English \\ \hline

Articles that explore challenges reviewers face when identifying security vulnerabilities during peer code review in OSS 
& Abstracts, proposals, posters, or short papers \\ \hline

Articles that examine factors that enhance reviewers’ ability to detect security vulnerabilities 
& Replication studies that do not provide new empirical evidence \\ \hline

\end{tabular}

%% file: tab_Search_DB.tex
\begin{tabular}{|l|c|c|}
\hline
\textbf{Database} & \textbf{May 2023} & \textbf{May 2023 -- Sept 2024} \\ \hline
IEEE Xplore & 420 & 21 \\ \hline
ACM & 63 & 14 \\ \hline
ScienceDirect & 80 & 25 \\ \hline
Springer & 189 & 44 \\ \hline
Scopus & 189 & 22 \\ \hline
\textbf{Total Retrieved} & \textbf{941} & \textbf{126} \\ \hline \hline
Duplicates & 9 & 2 \\ \hline
\textbf{Unique Results} & 932 & 124 \\ \hline
\end{tabular}

%% file: tab_Study_Selection.tex
\begin{tabular}{|l|c|c|}
\hline
\textbf{Selection stage} & \textbf{May 2023} & \textbf{May 2023 -- Sept 2024} \\ \hline
Unique search results & 932 & 124 \\ \hline
Title \& abstract & 416 & 24 \\ \hline
Introduction \& conclusion & 40 & 7 \\ \hline
Full-text & 15 & 5 \\ \hline
\end{tabular}

%% file: tab_Data_Extraction.tex
\begin{tabular}{|p{3cm}|p{8.75cm}|}
\hline
\centering\textbf{Data Item} &\textbf{Description}\\  \hline
Data Extractor  & Who extracts the data from the final set of papers\\ \hline
Data Checker & Who checks the extracted data to verify \\ \hline
Study Identifier& A unique digit identifier for a paper.  \\ \hline
Title & The title of the paper. \\ \hline
RQ1.  & What types of security vulnerabilities can be identified or missed during peer code review in OSS? \\ \hline
RQ2.  & What are the current security practices during peer code review in OSS?  \\ \hline
RQ3. & How do security analysis tools contribute to improving OSS security during peer code reviews? \\ \hline
RQ4. & What challenges and limitations do reviewers face in ensuring security during peer code reviews in OSS? \\ \hline
RQ5. & What factors enhance a reviewer’s ability to detect security vulnerabilities during peer code reviews in OSS? \\ \hline
\end{tabular}

%% file: sec-Result.tex
This section presents our findings based on the research questions. 
Throughout this section, we emphasize the key insights and discoveries. 
We also provide an in-depth analysis to contextualize these findings within the broader scope of our research objectives.
By examining each question thoroughly, we offer a comprehensive understanding of the implications and significance of the results.

\subsection{Classification of Security Issues}
\label{security_Classification}
Most primary studies included in this review report security issues using CWE identifiers or vulnerability types aligned with the CWE taxonomy.
In contrast, the Common Vulnerabilities and Exposures (CVE) database catalogs system-specific and publicly disclosed vulnerabilities after deployment. 
Therefore, CVE is less suitable for characterizing early-stage detection activities such as code review.

As discussed in \ref{sec_security_vul}, this review adopts the MITRE CWE framework to classify security issues identified or missed during peer code review.
The MITRE CWE taxonomy is organized hierarchically, grouping individual weakness types into progressively broader categories based on their underlying characteristics.
The taxonomy defines parent–child relationships among weakness types, allowing researchers to analyze vulnerabilities at different abstraction levels.
For example, CWE-79 (Cross-Site Scripting), CWE-77 (Command Injection), and CWE-89 (SQL Injection) are all subcategories of CWE-707 (Improper Neutralization).

To map the security issues reported in primary studies to high-level categories, we followed a bottom-up aggregation process using the CWE-1000 Research Concepts view~\cite{mitre_cwe_1000}.
For each CWE identifier reported in a primary study, we traversed the CWE hierarchy upward to identify its parent category, as described in the example above.
In addition, some primary studies reported security issues using informal vulnerability labels (e.g., ``injection flaws'' or ``XSS'') rather than explicit CWE IDs.
We matched these issues to the closest CWE entry based on the study's description.
This aggregation does not alter the original classification reported in the primary studies while enabling higher-level analysis of recurring security issue patterns.

Throughout this paper, we use the term vulnerabilities in a practical sense to refer to potential vulnerabilities, security weaknesses, flaws, or defects identified during code review, consistent with the terminology used in the primary studies. 
However, our analysis and classification are grounded in the CWE framework to ensure clarity and consistency.

\subsection{RQ1  What types of potential security vulnerabilities can be identified or missed during peer code review in OSS?}

We separate the discussion of the identified potential vulnerabilities from the discussion of those that were missed.

\subsubsection{High-level vulnerability types}

To simplify the analysis, we consolidated the identified and missed vulnerability categories into 10 high-level categories. 
We performed this reduction based on the hierarchical classification in the official CWE taxonomy, which defines relationships among different types of weaknesses. 
For example, CWE-79 (cross-site scripting), CWE-77 (command injection), and CWE-89 (SQL injection) are all subcategories under CWE-707 (Improper neutralization).
The following text explains each high-level category.

\paragraph{Improper Control of a Resource Through its Lifetime (CWE-664)} 
This vulnerability occurs when a system fails to properly manage a resource from its creation to its release. 
This category of vulnerabilities encompasses issues such as buffer overrun, information leakage, format string problem, improper synchronization, exposure of resource to wrong sphere, improper initialization, operation on a resource after expiration or release, and uncontrolled resource consumption, etc.

\paragraph{Improper Neutralization (CWE-707)}
This vulnerability occurs when a system fails to correctly handle, validate, or sanitize user input, allowing attackers to inject malicious data. 
This vulnerability can lead to injection-based vulnerabilities, such as cross-site scripting (XSS), cross-site request forgery (CSRF), SQL injection, and command injection. 

\paragraph{Improper Access Control (CWE-284)}
This vulnerability occurs when a system either fails to restrict or incorrectly restricts access to a resource from an unauthorized actor. 
This vulnerability type includes issues related to authentication, authorization, and accountability.

\paragraph{Improper Adherence to Coding Standards (CWE-710)}
This vulnerability occurs when a product fails to adhere to established coding rules or development standards.
This vulnerability type includes the use of potentially dangerous functions, third-party vulnerable packages, security misconfiguration, and documentation. 

\paragraph{Improper Check or Handling of Exceptional Conditions (CWE—703)}
This vulnerability occurs when a product fails to anticipate or handle exceptional conditions that occur during normal operation.
This vulnerability type includes crashes, catching all exceptions, and improper exception handling.

\paragraph{Insufficient Control Flow Management (CWE-691)}
This vulnerability occurs when code fails to properly control its execution path, allowing unexpected changes during program execution. 
This vulnerability type includes race conditions and uncontrolled recursion.

\paragraph{Protection Mechanism Failure (CWE-693)}
This vulnerability occurs when a system fails to use or improperly utilizes a protection mechanism to defend against attacks. 
This vulnerability type includes failing to protect network traffic and encryption. 

\paragraph{Incorrect Calculation (CWE-682)}
This vulnerability occurs when a system performs a calculation that generates incorrect or unintended results. 
This error may lead to issues such as incorrect resource allocations, improper privilege assignments, or inadequate resource management. 

\paragraph{Weakness that Affect System Process (CWE-634)}
This vulnerability occurs when defects in hardware and software negatively impact system-level operations. Note that, now this CWE category is listed as deprecated on cwe.mitre.org.

\paragraph{C++ Catastrophes} 
This vulnerability does not have an assigned CWE identifier, as it refers to critical problems specific to the C++ language that can lead to severe security flaws.

\subsubsection{Identified vulnerabilities}

The analysis of RQ1 shows that peer reviews in OSS are capable of uncovering a wide range of vulnerabilities, which we organized into high-level categories based on the CWE classification. 
Across the 20 primary studies, we identified 46 distinct vulnerability types detected during code review, with a total reported frequency of 1,789 instances spanning ten high-level CWE categories.

The most prevalent category was \textit{Improper Control of a Resource Through its Lifetime (CWE-664)}, accounting for 587 instances.
Common vulnerabilities under this category included buffer overrun, improper synchronization, improper resource shutdown or release, and resource leak.
The second most prevalent category was \textit{Insufficient Control Flow Management (CWE-691)}, with 464 instances, primarily involving race conditions and uncontrolled recursion.
The third most prevalent category was \textit{Improper Check or Handling of Exceptional Conditions (CWE-703)}, comprising 203 instances, most often related to crashes, improper checks, and catching all exceptions. 
Moderately frequent categories included \textit{Improper Access Control (CWE-284)}, which was reported in 152 instances, covering vulnerabilities such as improper authentication and too much privileges.
\textit{Incorrect Calculation (CWE-682)} was found in 151 cases, typically integer overflows and incorrect calculations.
\textit{Improper Neutralization (CWE-707)} was also notable, with 126 cases, including client-side cross-site scripting (XSS), SQL injection, code injection, and improper input validation. 
\textit{Improper Adherence to Coding Standards (CWE-710)} was noted in 71 instances, encompassing the use of potentially dangerous functions, inadequate documentation, and the use of vulnerable packages. 
The less frequently identified categories were \textit{Protection Mechanism Failures (CWE-693)}, with 26 instances, mainly encryption weaknesses.
A specialized category, the language-specific \textit {C++ Catastrophe}, was observed in 8 instances, and \textit{Weaknesses Affecting System processes (CWE-634)} appeared in just one instance.

A complete list of identified vulnerabilities and their frequencies is provided in Table \ref{tab:vulnfound} in the Appendix section.

\subsubsection{Missed Vulnerabilities}

Our findings reveal that peer code review in OSS frequently overlooks a diverse set of security vulnerabilities.
Across the 20 primary studies, we identified 38 distinct vulnerability types that were missed during peer code review, with a total reported frequency of 461 instances spanning the same ten high-level CWE categories.

The most frequently missed category was \textit{Improper Control of a Resource through its Lifetime (CWE-664)} with 180 instances.
The vulnerabilities under this category were operation on a resource after expiration or release, and exposure of the resource to the wrong sphere.
The second most common category was \textit{Improper Neutralization (CWE-707)}, comprising 102 instances. 
The most commonly overlooked vulnerabilities in this category were improper input validation, client-side cross-site scripting (XSS), and data-processing errors.
The third most frequently overlooked category is \textit{Improper Access Control (CWE-284)}, with 65 instances, primarily involving general improper access control and too much privileges.
The special category of \textit{C++ Catastrophe} vulnerabilities accounted for 33 missed cases. \textit{Improper check or handling of exceptional conditions (CWE-703)} was missed in 22 instances, while \textit{incorrect calculation (CWE-682)} occurred in 20 instances.
Both \textit{Insufficient Control Flow Management (CWE-691)} and \textit{Protection Mechanism Failure (CWE-693)} accounted for 17  missed instances.
The former was predominantly associated with race conditions, while the latter was related to insufficient verification of data authenticity.
\textit{Improper Adherence to Coding Standards (CWE-710)} appeared in 4 instances, linked to the use of potentially dangerous functions.
Finally, \textit{Weaknesses Affecting System processes (CWE-634)} was the least frequent, with only one instance recorded.

A complete list of identified vulnerabilities and their frequencies is provided in Table \ref{tab:missedvuln} in the Appendix section.

\subsection{RQ2	What are the current security practices during peer code review in OSS?}

We grouped the identified practices into the categories described below.

\subsubsection{Reviewer Involvement}

The involvement of multiple reviewers in peer code reviews is a common practice that can significantly enhance the detection of security vulnerabilities in OSS, particularly when reviewers have relevant security knowledge~\cite{dibiaseatal2016, mcintosh2014}. 
Code reviews in the Chromium open-source project found that reviews with more than two reviewers were more effective in identifying security vulnerabilities~\cite{dibiaseatal2016}. 
Similarly, prior studies~\cite{Thongtanunametal2015, kononenkoetal2015} have observed that a smaller number of reviewers is associated with a higher risk of overlooking security vulnerabilities.
However, the relationship between the number of reviewers and security outcomes is not consistently supported across the literature. 
Meneely et al.~\cite{meneelyatel2014} observed that vulnerable files often had more reviewers, many of whom lacked security knowledge. 
Paul et al.~\cite{rajpauletal2021} found reviewers who perform an excessive number of reviews may become less caution and miss security vulnerabilities.
Thompson et al~\cite{thompsonetal2017} found no statistically significant relationship between the number of reviewers and the identification of security vulnerabilities.
These findings suggest that an optimal number of reviewers, balanced with security knowledge, is critical for achieving effective security outcomes.

\subsubsection{Reviewer Expertise and Code Familiarity}

Reviewer expertise, familiarity with the codebase, and project context have a significant impact on vulnerability detection in open-source software. 
Reviews conducted by developers with prior experience in identifying or fixing vulnerabilities are more likely to detect critical vulnerabilities~\cite{Davilaetal2021, meneelyatel2014}.
In addition, increased reviewer expertise and closer collaboration between reviewers and authors reduces the likelihood of post-release security vulnerabilities~\cite{meneelyatel2014}. 
In contrast, reviews involving participants with limited security knowledge and unfamiliarity with the relevant code are more likely to miss vulnerabilities~\cite{mcintosh2014, kononenkoetal2015}. 

However, the relationship between experience and vulnerability detection is not entirely straightforward. 
One study reported that highly experienced developers were often responsible for introducing vulnerable code~\cite{carverandbosu2014}.
However, code changes by less experienced authors are significantly more likely to be vulnerable, ranging from 1.5 to 24 times higher than those by experienced authors~\cite{carverandbosu2014}. 
Another study found no clear relationship between the reviewer's level of experience and detection accuracy, suggesting that specific security knowledge may outweigh general experience~\cite{edmundsonetal2013}.
These findings suggest that while reviewer expertise and project familiarity are essential for detecting security vulnerabilities during code review, experience alone is insufficient. 

\subsubsection{Code Change Characteristics}

The size and complexity of code changes influence the likelihood of security issues being detected during review. 
Smaller changes and clearly articulated review requests tend to draw greater reviewer participation and scrutiny~\cite{Davilaetal2021}. 
In contrast, larger code modifications and high code churn increase the risk of introducing vulnerabilities, as their complexity can overwhelm reviewers~\cite{carverandbosu2014, meneelyatel2014}. 
Additionally, modified files are more prone to vulnerabilities than new ones, necessitating heightened scrutiny for alterations to existing code~\cite{carverandbosu2014}. 
These insights suggest that OSS projects can enhance security by promoting concise and well-documented code changes.

\subsubsection{Addressing vulnerabilities}

While peer code review is a widely adopted practice for improving software quality and identifying potential vulnerabilities, empirical evidence suggests that vulnerability detection often receives less emphasis than expected during the review process.
Bacchelli and Bird~\cite{bacchbird2013} found that code reviews tend to focus more on general quality concerns than on security-specific issues. 
Charoenwet et al.\cite{wachiraphanandel2024} further support this pattern by showing that developers identified and discussed coding weaknesses significantly more frequently than security vulnerabilities, by a factor ranging from 21 to 33.5. 
These findings indicate that although security is a recognized objective, it is frequently overshadowed by other review priorities.

When code review identifies security issues, the developers' responses vary considerably. 
One study~\cite{alfadelandandnagy2023} reported that 54.96\% of security-related concerns raised during code review were addressed through fixes or mitigations. 
Nonetheless, a substantial proportion of issues (28.65\%) were assessed as low risk and not prioritized for immediate resolution.
Additionally, a notable percentage of issues remained unresolved (8.18\%) or received no response at all (4.67\%). 
Similarly, another study~\cite{wachiraphanandel2024} found that while developers often attempted to resolve identified concerns within the same pull request, 30\% to 36\% of these issues were acknowledged and merged without resolution, and 15\% to 26\% were either dismissed or left unaddressed.

The degree of reviewer engagement in guiding the resolution of security issues also varies. 
According to one study~\cite{yuandfu2023}, over half of the reviewers (53.8\%) anticipated that developers would take responsibility for addressing the reported vulnerabilities.
In most of these instances (86.6\%), the reviewers provided explicit guidance, including suggested fixes or code snippets. 
However, in 13.4\% of the cases, reviewers identified issues without offering actionable advice. 
Furthermore, in 39\% of the cases, reviewers noted the presence of a vulnerability without indicating any next steps, and in 7.2\% of cases, reviewers explicitly recommended ignoring the issue, often citing its perceived insignificance.
Collectively, these findings highlight inconsistencies in how security concerns are addressed during code review. 
These patterns suggest a need for more structured practices and clearer expectations to support the consistent handling of security vulnerabilities in the peer review process.

\subsubsection{Human and Organizational Factors}

Human and organizational dynamics have a significant impact on security practices in OSS code reviews.
A study by Braz et al.~\cite{brazandbacchelli2022} provided the following insights.
Developers recognize the importance of security but often give it a lower priority unless explicitly prompted otherwise, and sometimes assume their applications are inherently secure. 
Additionally, the majority of organizations fail to provide adequate security training, yet they expect developers to address security issues during code reviews.
Moreover, developers’ efforts in security contributions are often unrecognized, which may further reduce the focus on security during code reviews. 
Another study~\cite{brazandchristian2022} demonstrated that explicitly instructing reviewers to focus on security during code review increased the likelihood of vulnerability detection by a factor of eight. 
This insight suggests that without support from the organization and cultural emphasis on security, even skilled reviewers and good practices may not be enough. 
It highlights the need for better training and recognition in open-source projects.

\subsection{RQ3 How do security analysis tools contribute to improving OSS security during peer code reviews?}
\label{sec_tools}
Security analysis tools, particularly static application security testing (SAST) tools, significantly enhance the security of open-source software (OSS) during code reviews. 
These tools assist reviewers by flagging potentially vulnerable code segments, thereby reducing the manual effort required from reviewers and allowing them to focus on more complex aspects of the code.
One study~\cite{charoenwet2024} demonstrated that prioritizing functions based on SAST warnings improves the identification of vulnerabilities, leading to a 12\% increase in precision, a 5.6\% increase in recall, and a 13\% reduction in false positives compared to non-prioritized reviews.
This result suggests that integrating the output of security analysis tools into review workflows significantly reduces the burden on human reviewers and improves the overall quality of the review process.

The integration of security tools into development workflows further supports early detection of common vulnerabilities. 
For instance, individual SAST tools were capable of detecting approximately 52\% of known vulnerabilities in modified functions, and combining multiple tools increased detection rates by an additional 26\%~\cite{charoenwet2024}. 
In addition, several projects have adopted automated tools in continuous integration (CI) pipelines to support security checks during the review process~\cite{alfadelandandnagy2023}.
One such example is the Marked project in the npm ecosystem, which uses the vuln-regex-detector tool to identify ReDOS vulnerabilities~\cite{davisjam2021vulnregex}. 
These cases demonstrate how the integration of security tools can reliably identify security issues early within the development workflow.

Despite these benefits, security analysis tools have notable limitations. 
The evaluated SAST tools generated a large number of irrelevant warnings, with 76\% being false positives. 
Furthermore, they failed to produce any warning for 22\% of vulnerable code changes (VCCs)~\cite{charoenwet2024}.
These results highlight the significant limitations of current SAST tools.
Therefore, the effective use of static analysis tools requires selecting tools that are aligned with the project’s specific security requirements and development context, as well as supplementing the tool output with human judgment and manual review.

\subsection{RQ4 What challenges and limitations do reviewers face in ensuring security during peer code reviews in OSS?}

Peer code reviews play a vital role in improving security in OSS.
However, reviewers face several challenges that limit their effectiveness. 
This section outlines the key challenges identified in prior studies.

\subsubsection{Insufficient Security Knowledge and Expertise}

A significant challenge to ensuring security during code reviews in OSS is the reviewers' insufficient security knowledge and lack of secure coding practices. 
Prior research consistently highlights that developers often lack security knowledge, secure coding practices, and have limited familiarity with common vulnerability patterns~\cite{carverandbosu2014, meneelyatel2014, brazandbacchelli2022}. 
One study~\cite{carverandbosu2014} found that different developers introduce similar types of security vulnerabilities, such as unsafe method usage, failure to sanitize inputs, TOCTTOU race conditions, and integer arithmetic overflows, across code changes. 
These vulnerabilities can be mitigated by acquiring security knowledge and adopting secure coding practices. 
Insufficient security training is also a significant challenge in effectively ensuring security during code reviews~\cite{brazandbacchelli2022}. 
Furthermore, reviewers frequently encounter challenges in ensuring security, particularly when they have limited security experience and are unfamiliar with collaborators and the projects~\cite{kononenkoetal2015, mcintosh2014, meneelyatel2014}.

\subsubsection{Time Constraints and Review Fatigue}

Time constraints and high review volumes significantly impact reviewers' ability to focus on security. 
Time constraints are particularly prevalent, leading to superficial review efforts and reduced attention to security considerations~\cite{brazandbacchelli2022, rajpauletal2021}.
In addition, high review volumes lead to review fatigue, thereby reducing the likelihood of detecting security vulnerabilities~\cite{rajpauletal2021}.
Reviewing a high volume of code under time constraints can lead to review fatigue, which reduces a reviewer’s ability to effectively identify security vulnerabilities.

\subsubsection{Contextual and Practical Constraints}

Reviewers face several contextual and practical constraints that limit their ability to identify security vulnerabilities.
Code changes that affect multiple directories and files significantly expand the review scope, making it more difficult for reviewers to detect security vulnerabilities~\cite{rajpauletal2021}.
Many reviewers lack a comprehensive understanding of the application under review, including the behavior of its components, third-party library dependencies~\cite{brazandbacchelli2022, bacchbird2013}.
These challenges make it difficult to assess whether a given code change might expose the system to new threats. 
In addition, the broad and evolving landscape of potential vulnerabilities, combined with diverse and subtle attack patterns, further complicates the identification of security vulnerabilities during code reviews~\cite{brazandbacchelli2022}. 

\subsubsection{Human Factors and Process Limitations}

Human factors, combined with inherent limitations in the code review process, make it more challenging for reviewers to identify vulnerabilities. 
For example, some vulnerabilities only become apparent during code execution, making them difficult to spot through manual peer code review alone~\cite{brazandbacchelli2022}. 
Human factors, such as limited security awareness, reviewers' mindset, and cognitive load, also hinder the effectiveness of security-focused code reviews. 
Reviewers may not prioritize or recognize the importance of security concerns unless explicitly directed to do so~\cite{brazandbacchelli2022, bacchbird2013}. 

\subsection{RQ5 What factors enhance a reviewer’s ability to detect potential security vulnerabilities during peer code reviews in OSS? }

Understanding the factors that enhance a reviewer’s ability to detect security vulnerabilities is essential for improving the effectiveness of peer code reviews in OSS. 
This section presents the key factors identified in the literature.

\subsubsection{Security Knowledge and Expertise}

The reviewer’s security knowledge strongly influences their ability to detect security vulnerabilities during code review. 
One study~\cite{brazandbacchelli2022} found that reviewers with security knowledge acquired through formal education, security training, or participation in workshops demonstrate greater proficiency in identifying vulnerabilities during code reviews. 
Security expertise is a critical factor in effective vulnerability detection during code review. 
Reviewers with a strong foundation in security are better equipped to identify potential vulnerabilities~\cite{Davilaetal2021, meneelyatel2014}. 
This expertise encompasses both broad knowledge of security principles and specific experience in vulnerability identification~\cite{meneelyatel2014}.

\subsubsection{Review Background and Codebase Familiarity}
\label{sec_reviewer-background}
Familiarity with the codebase and prior review experience significantly enhances a reviewer’s ability to detect security vulnerabilities. 
Reviewers with previous experience reviewing a specific file or with a history of collaboration with the code author are more likely to identify security issues~\cite{rajpauletal2021, meneelyatel2014}.
Familiarity with the codebase and active engagement in the review process help reviewers better understand the context and impact of code changes, which significantly improves their ability to detect security vulnerabilities~\cite{Davilaetal2021, meneelyatel2014}.

\subsubsection{Mental Attitude and Security Awareness}

The reviewer’s mental attitude and focus toward security practices also shape their effectiveness in identifying vulnerabilities.
One study~\cite{brazandchristian2022} found that developers who are explicitly instructed to prioritize security concerns during code review are more successful in identifying vulnerabilities compared to those who are not.
In addition, their results found that the mental attitude of reviewers and consistent adherence to secure coding guidelines play a significant role in detecting security vulnerabilities during code reviews.
Similarly, another study~\cite{brazandbacchelli2022} further emphasizes that organizational culture and guidelines that promote security awareness can reinforce this mindset, enhancing reviewers’ vigilance and reducing the likelihood of overlooking common vulnerabilities.

\subsubsection{Training, Support, and Review Environment}
\label{sec_organizationa_aspect}
Organizational and contextual factors substantially influence the effectiveness of security-focused code reviews.  
One study~\cite{brazandbacchelli2022} emphasizes that participation in security-specific training or workshops enhances reviewers’ ability to identify vulnerabilities by providing relevant knowledge and up-to-date techniques. 
The same study showed that access to security expert support, resources, and rewards for vulnerability detection can motivate reviewers to evaluate code with a security-oriented perspective.
These findings underscore the importance of a structured and supportive review environment, characterized by formal training opportunities, expert guidance, reward, and adequate time allocation to enhance the quality and reliability of security assessments during code reviews.

\subsubsection{Review Duration and Engagement}

The amount of time spent on a code review and the reviewer’s level of engagement play an essential role in identifying security vulnerabilities. 
Longer review sessions provide reviewers with more opportunities to examine the code in detail, thereby increasing the likelihood of identifying security vulnerabilities~\cite{rajpauletal2021}.
However, extended reviews can lead to review fatigue, which may reduce effectiveness.
In addition to time, the quality of reviewer engagement is also essential.
Reviewers who are actively involved and thorough in their assessments are more likely to detect security vulnerabilities~\cite{Davilaetal2021}.
These findings suggest that sufficient time and focused effort are essential to conducting effective security-oriented code reviews.

%% file: sec-Discussion.tex
This section synthesizes the findings from our systematic literature review to provide a comprehensive interpretation of the results.
This discussion also outlines practical implications and future directions to enhance the effectiveness of peer code reviews as a security assurance mechanism.
In addition, recent advances in AI-assisted code review introduce new opportunities and challenges. We discuss these emerging trends in Section 5.6.

\subsection{RQ1 What types of potential security vulnerabilities can be identified or missed during peer code review in OSS?}

The findings from RQ1 reveal clear patterns in reviewers’ strengths and limitations when identifying security vulnerabilities during OSS peer code reviews. 
Reviewers are generally effective at detecting vulnerabilities that are simple, well-known, visible, and localized in the code change.
Examples include buffer overruns, race conditions, crashes, XSS, and the use of well-known dangerous functions.
These vulnerabilities are widely documented in secure coding guidelines that align closely with developers’ programming knowledge and are easier to spot through code review.
Conversely, vulnerabilities that are complex, subtle, semantic, context-dependent, and necessitate code execution or a thorough understanding of code context, such as improper input validation, improper access control, and resource lifecycle errors, were less likely to be detected. 
These issues often require deeper reasoning about broader system behavior, data propagation, and privilege boundaries, which are difficult to infer from isolated code snippets.
These findings align with the results of the prior study by Paul et al~\cite{rajpauletal2021}, which highlight the limitations of manual reviews for semantic or intricate flaws.

To enhance effectiveness, OSS communities could prioritize security knowledge and training, structured review checklists, and integration of automated tools to complement human oversight. 
Such measures can help reduce overlooked vulnerabilities and strengthen the overall security of OSS projects.

\subsection{RQ2 What are the current security practices during peer code review in OSS?}
\label{sec_organizational_dis}
The results for RQ2 indicate that effective security practices in OSS peer code reviews depend on the alignment of technical practices and human–organizational supports, including reviewer involvement, expertise, security knowledge, code familiarity, code change characteristics, and managerial support. 
Reviews are most effective when they involve reviewers who not only possess security expertise but are also familiar with the codebase and project context. 
These reviewers are better able to interpret code changes accurately and identify security issues, especially when the changes are small, well-documented, and clearly scoped.
Such conditions help focus the reviewer's effort and reduce the risk of overlooking vulnerabilities.

However, the results also reveal essential gaps and inconsistencies in current practices.
Code quality or coding style concerns often overshadow security issues during code reviews.
Even when reviewers identify vulnerabilities, their responses are inconsistent. 
Reviewers often classify many vulnerabilities as low risk, leave them unresolved, or do not follow up correctly during the review process. 
This approach to handling vulnerabilities during reviews reflects a lack of structured processes for prioritizing and resolving security issues.
Moreover, human and organizational factors have a significant impact on security practices in OSS code reviews.
Developers often deprioritize security unless explicitly asked to focus on it.
Many organizations expect reviewers to address security, but do not provide training or recognition for doing so. 

Overall, these patterns highlight the need for OSS communities to foster a security-centric culture. 
This need includes balancing the number and expertise of reviewers, ensure project and code familiarity, managing code change size, structured protocols for prioritizing and resolving security issues, providing security training, and promoting a security oriented mindset.

\subsection{RQ3 How do security analysis tools contribute to improving OSS security during peer code reviews?}

The integration of security analysis tools, especially SAST, plays a key role in supporting the identification of security issues in OSS during peer code reviews.
These tools help reduce the cognitive load on human reviewers by automatically flagging potentially vulnerable code patterns.
This automated detection allows reviewers to focus their attention on more complex or contextual issues. 
However, the results also reveal that tool-based approaches alone are insufficient.
Tools often miss specific vulnerabilities, produce high false positives, and lead to reviewer overload due to excessive false alerts.
These limitations underscore the importance of carefully selecting tools that align with a project’s security requirements. 
Overall, these findings highlight that security analysis tools are most effective when used in conjunction with a human reviewer. 
Embedding tools in review workflows can improve efficiency and coverage, but their success ultimately relies on knowledgeable reviewers who can contextualize tool findings and make informed security decisions.

\subsection{RQ4 What challenges and limitations do reviewers face in ensuring security during peer code reviews in OSS?}

The findings for RQ4 reveal that reviewers face several interconnected challenges that hinder their ability to ensure security during OSS code reviews. 
The most significant challenge is the reviewers' limited knowledge and expertise in security.
Empirical studies demonstrate that many reviewers lack sufficient familiarity with secure coding practices and common vulnerability patterns, which increases their likelihood of overlooking critical security issues.
Limited security training opportunities and unfamiliarity with project-specific contexts and collaborators worsen the problems.
In addition, review fatigue and time constraints often force reviewers to prioritize functionality and coding style over security issues.

Human and organizational factors, as well as contextual and process limitations, further complicate the reviewers' ability to identify security vulnerabilities.
Reviews become more complex when code changes span multiple directories, involve third-party libraries, or modify large subsystems.
Human and cultural factors, such as reviewer mindset, limited awareness, and high cognitive load, also reduce their likelihood of detecting subtle vulnerabilities.
Moreover, reviewers may not prioritize security concerns unless they are explicitly instructed to do so, reflecting a broader cultural tendency to treat security as secondary to other review goals.

In summary, reviewers face challenges that include a lack of security knowledge, human and cognitive limitations, organizational shortcomings, and practical constraints within the review process. 
Addressing these challenges requires security focused training to enhance reviewer knowledge, structured review practices to manage workload effectively, the integration of automated tools to complement human judgment, and a stronger organizational emphasis on security during code reviews.

\subsection{RQ5 What factors enhance a reviewer’s ability to detect potential security vulnerabilities during peer code reviews in OSS?}

The findings for RQ5 show that a reviewer’s ability to detect security vulnerabilities in OSS depends on both technical skills and the environment in which reviews take place. 
Reviewers who possess strong security knowledge and experience and have attended security training and workshops are more effective at spotting vulnerabilities.
Familiarity with the project and collaborators is also essential.
Reviewers who understand the context and history of the code are better able to identify security issues.
In addition, a security-focused attitude and awareness significantly affect security outcomes. 
Reviewers who approach code review with an explicit focus on security or are prompted by organizational guidelines to prioritize security concerns are more likely to identify vulnerabilities.

Organizational and contextual factors further enhance reviewer effectiveness. 
Access to security training, support from security experts, and recognition or rewards for finding vulnerabilities all motivate reviewers to take security issues more seriously.
In addition, spending sufficient time and maintaining focus during code reviews enables reviewers to examine changes in depth, thereby enhancing their ability to detect security vulnerabilities. 
However, excessive review duration can lead to review fatigue and negatively impact performance.

In summary, a reviewer's ability to detect security vulnerabilities in OSS depends on security knowledge and experience, codebase familiarity, security-focused attitudes and awareness, organizational support, and sustainable review practices. 
Future research could investigate these factors in various OSS settings and examine how combining reviewer expertise with AI-based support impacts their effectiveness.

\subsection{Emerging Trends: AI-Augmented Code Review for Security in OSS}

This review focuses on manual peer code review rather than automated or AI-powered code review.
However, recent advances in generative AI, particularly large language models (LLMs), have opened new opportunities to support software engineering tasks.
LLMs have demonstrated strong capabilities in code understanding, including code generation, summarization, vulnerability pattern recognition, program repair, and assistance with static analysis~\cite {brownetal2020, longetal2022, xiaetal2023, lietal2024}.
These capabilities have motivated growing interest in AI-assisted techniques that augment traditional peer code-review workflows.

From a security perspective, AI-based approaches can address several limitations identified in this SLR. 
For example, our findings in RQ1 show that reviewers often miss vulnerabilities that require contextual reasoning and broader execution context. 
Similarly, RQ4 highlights the impact of limited expertise, time pressure, and cognitive load on vulnerability detection. 
AI-assisted analysis can help mitigate these challenges by automatically identifying suspicious patterns, prioritizing high-risk code regions, and reducing reviewer workload.
In this way, AI support aligns with the needs identified in this study, particularly in assisting reviewers with limited security expertise.

Despite these opportunities, current evidence suggests that AI techniques are not yet reliable as standalone solutions. 
Recent empirical studies indicate that general-purpose LLMs struggle with deep semantic reasoning, data-flow analysis, and project-level context~\cite{steenhoeketal2024, khareetal2024}.
Moreover, LLMs may generate hallucinated or verbose outputs, which raises concerns about trust, explainability, and reliability in security-critical environments~\cite{xuetal20252025}. 
Privacy, bias, and ethical considerations also limit the direct use of these methods in OSS settings, where code may be sensitive or confidential.

Overall, these findings reinforce that AI-assisted tools can be viewed as a complementary aid in security-focused code review, but they cannot replace human judgment.
AI-assisted tools can support vulnerability detection by guiding attention and improving prioritization, while human reviewers remain essential for reasoning and security decision-making.
Future research should empirically evaluate the impact of AI-assisted review in real OSS repositories and explore hybrid human–AI workflows that improve both effectiveness and efficiency.

\subsection{Implications of OSS Development Context}

The findings of this review highlight several factors that affect the effectiveness of security issues detection during code review. 
However, the role of the OSS development context itself remains underexplored in the existing literature. 
OSS projects are often characterized by geographically distributed contributors, asynchronous collaboration \cite{Xu2025,Guizani2021,Palazzi2019}, varying levels of expertise \cite{Trinkenreich2020}, and diverse governance and sponsorship models \cite{Linaker2025,Osborne2024}.
These characteristics may influence reviewers’ ability to understand system architecture, project history, and design rationale. 
As discussed in Section~\ref{sec_reviewer-background}, familiarity with the codebase is a key factor in detecting security issues. 
However, the included studies rarely examine how these factors are shaped by OSS-specific development environments.

The distributed and asynchronous nature of OSS development also has implications for tool support.
As discussed in Section~\ref{sec_tools}, security analysis tools can provide a baseline guidance across diverse contributors.
However, many studies evaluated these tools using OSS repositories mainly as empirical testbeds rather than explicitly analyzing how OSS-specific characteristics influence tool effectiveness.

Similarly, organizational and governance aspects discussed in Sections~\ref{sec_organizationa_aspect} and ~\ref{sec_organizational_dis} may differ across OSS settings. 
Community-driven projects often rely on informal guidelines and experienced maintainers, whereas sponsored OSS projects may provide structured policies, training, rewards and incentives, and security expertise.
The included studies rarely distinguish between these governance models. 
Therefore, future research should investigate how governance structure, contributor distribution, and collaboration practices influence security-focused code review in real-world OSS environments.

%% file: sec-Threat_to_validity_new.tex
This section describes the threats to validity of the study and how we addressed them, where possible.

\subsection{Search Strategy and Study Selection}

The quality and coverage of an SLR depend heavily on the search strategy and study selection process. 
To identify relevant studies, we designed the search string using terms derived from prior work on software security, peer code review, and OSS development.
We refined the search string through pilot searches and discussions between the authors to improve coverage and reduce bias.

We searched multiple digital libraries covering software engineering journals, conferences, and workshops. 
Given that there is no single standard venue for publishing this type of work, some relevant studies may still have been missed.
However, we additionally performed backward snowballing to mitigate this threat as much as possible.

We also used predefined inclusion and exclusion criteria during study selection.
Two authors independently screened a subset of studies at different stages of the review process and resolved disagreements through discussion.
Despite these efforts, some relevant studies may still have been missed due to differences in terminology, database indexing limitations, or publication availability.

\subsection{Security Vulnerability Classification}

Most primary studies reported potential vulnerabilities or security issues using CWE identifiers.
To maintain consistency across studies, we mapped the reported vulnerabilities to high-level CWE categories using the official MITRE taxonomy~\cite{mitre_cwe}.
However, some studies described vulnerabilities only by name and did not explicitly provide CWE identifiers.
In those cases, the mapping process required interpretation, which may introduce classification bias.
To reduce this risk, both authors reviewed the mappings and resolved disagreements through discussion. 
Nevertheless, some degree of subjectivity may remain in the classification of vulnerabilities across studies. 

\subsection{Scope and Generalizability}

This review focuses on peer code review in OSS projects.
Many included studies examined large and well-established OSS projects such as Chromium, Qt, OpenStack, Mozilla, and npm packages. 
Therefore, the findings may not generalize to smaller community-driven OSS projects or proprietary software environments.
In addition, OSS projects differ in governance structure, contributor expertise, sponsorship, and development practices.
Most primary studies did not explicitly distinguish among these contextual factors, which may limit the applicability of the findings across different OSS ecosystems.

\subsection{Reliability}

Reliability concerns whether the review process is sufficiently transparent and consistent for another research team to follow the same procedure.
To improve consistency and reliability, we used structured data extraction forms and systematically synthesized findings across studies.
Two authors participated in the study selection and vulnerability classification process and resolved disagreements through discussion. 
The documented review protocol, independent screening process, and transparent reporting of decisions improve the replicability of the review and increase confidence in the stability of the findings.

\vspace{8pt}

%% file: sec-Conclusion.tex
This paper presents the first SLR on the role of peer code review in identifying security vulnerabilities in OSS. 
While peer code review has long been recognized as a valuable practice for improving code quality, studies of its effects on security are relatively recent.
By synthesizing findings from 20 primary studies, this review provides a comprehensive overview of the vulnerabilities identified and missed during code reviews, the practices and tools that support security detection, and the challenges and factors influencing reviewer effectiveness.

The results show that peer code reviews are effective in detecting common, localized vulnerabilities, particularly those that require understanding only a few lines of code. 
However, reviewers frequently overlook more complex vulnerabilities that require reasoning about execution context, resource lifecycles, or system-level interactions.
Security practices emphasize balancing the number of reviewers, their expertise, project familiarity, and prioritizing high-risk changes.
However, inconsistencies in addressing identified security concerns highlight gaps in security practices in the code review process. 

Security analysis tools, particularly SAST integrated into continuous integration (CI) pipelines enhance vulnerability detection. 
Their effectiveness is maximized when used in combination with human judgment that suggests the importance of hybrid human–tool approaches.

Reviewers face challenges such as insufficient security knowledge, training, limited expertise, time pressure, and difficulties understanding complex systems and third-party dependencies.
Human and organizational factors like security awareness, lack of explicit guidance, and cultural tendencies further contribute to the low prioritization of security during code reviews.
The factors that influence reviewer effectiveness depend on both technical and organizational conditions, including security knowledge, attitudes, familiarity with the code base, training, structured review practices, and organizational support.

The peer code review process alone cannot guarantee complete security.
However, when combined with structured checklists, automated analysis tools, and a security-focused review culture, proper acknowledgment of effort can substantially reduce post-release vulnerabilities in OSS. 
For researchers, this SLR establishes a baseline for evaluating the security impact of peer code reviews, bridging gaps between code quality and vulnerability mitigation. 
Future research should focus on developing and integrating AI-assisted security analysis tools into code review workflows, reviewer-centered guidelines, and exploring approaches to reduce the challenges that reviewers face during code reviews.

%% file: sec-Appendix.tex
\begin{landscape}
\centering
\small
\setlength{\tabcolsep}{4pt}
\begin{longtable}{|m{5.9cm}|p{7cm}|p{3.5cm}|p{1cm}|}
\caption{Identified potential security vulnerabilities}
\label{tab:vulnfound}\\
\hline
\textbf{CWE Category (High level)} & \textbf{Vulnerability Name} & \textbf{Citations} & \textbf{Freq.} \\ \hline
\endfirsthead

\hline
\textbf{CWE Category (High level)} & \textbf{Vulnerability Name} & \textbf{Citations} & \textbf{Freq.} \\ \hline
\endhead

\endfoot

\hline
\endlastfoot

\multirow{20}{6cm}{\textbf{Improper Control of a Resource Through its Lifetime (CWE-664)}}
 & Buffer Overruns & \cite{dibiaseatal2016,nagyandal2009,alfadelandandnagy2023,carverandbosu2014,yuandfu2023,Anetal2018} & 118 \\ \cline{2-4}
 & Information Leakage & \cite{dibiaseatal2016,carverandbosu2014,wachiraphanandel2024,mahmoudetal2023} & 16 \\ \cline{2-4}
 & Format String Problems & \cite{dibiaseatal2016,nagyandal2009,carverandbosu2014,yuandfu2023} & 15 \\ \cline{2-4}
 & Use of Magic URLs / Predictable Cookies & \cite{dibiaseatal2016} & 1 \\ \cline{2-4}
 & Incorrect Type Conversion or Cast & \cite{dibiaseatal2016,rajpauletal2021} & 7 \\ \cline{2-4}
 & Improper Control of a Resource in its Lifetime & \cite{dibiaseatal2016} & 1 \\ \cline{2-4}
 & Regular Expression DoS (ReDoS) & \cite{alfadelandandnagy2023,mahmoudetal2023} & 23 \\ \cline{2-4}
 & Sensitive Data Exposure & \cite{alfadelandandnagy2023} & 10 \\ \cline{2-4}
 & Deadlock & \cite{alfadelandandnagy2023,carverandbosu2014,yuandfu2023} & 31 \\ \cline{2-4}
 & Denial of Service (DoS) & \cite{alfadelandandnagy2023,carverandbosu2014,yuandfu2023,mahmoudetal2023} & 33 \\ \cline{2-4}
 & Memory Leak & \cite{carverandbosu2014,Anetal2018} & 2 \\ \cline{2-4}
 & Resource Leak & \cite{yuandfu2023} & 59 \\ \cline{2-4}
 & Use After Free & \cite{yuandfu2023} & 8 \\ \cline{2-4}
 & Improper Synchronization & \cite{rajpauletal2021} & 109 \\ \cline{2-4}
 & Exposure of Resource to Wrong Sphere & \cite{rajpauletal2021} & 15 \\ \cline{2-4}
 & Improper Initialization & \cite{rajpauletal2021} & 5 \\ \cline{2-4}
 & Improper Resource Shutdown or Release & \cite{rajpauletal2021} & 71 \\ \cline{2-4}
 & Operation on a Resource after Expiration or Release & \cite{rajpauletal2021} & 3 \\ \cline{2-4}
 & Improper Restriction of Operations within Bounds of Memory Buffer & \cite{rajpauletal2021,wachiraphanandel2024} & 53 \\ \cline{2-4}
 & Uncontrolled Resource Consumption & \cite{wachiraphanandel2024} & 7 \\ \hline

\multirow{2}{6cm}{\textbf{Insufficient Control Flow Management (CWE-691)}}
 & Race Conditions & \cite{dibiaseatal2016,alfadelandandnagy2023,carverandbosu2014,yuandfu2023,rajpauletal2021,wachiraphanandel2024} & 456 \\ \cline{2-4}
 & Uncontrolled Recursion & \cite{wachiraphanandel2024} & 8 \\ \hline

\multirow{6}{6cm}{\textbf{Improper Neutralization (CWE-707)}} 
 & Client-side XSS & \cite{dibiaseatal2016,alfadelandandnagy2023,carverandbosu2014,yuandfu2023,edmundsonetal2013,mahmoudetal2023} & 72 \\ \cline{2-4}
 & SQL Injection & \cite{alfadelandandnagy2023,carverandbosu2014,yuandfu2023,edmundsonetal2013} & 27 \\ \cline{2-4}
 & Remote Code Injection & \cite{alfadelandandnagy2023,mahmoudetal2023} & 9 \\ \cline{2-4}
 & Improper Input Validation & \cite{alfadelandandnagy2023,rajpauletal2021,mahmoudetal2023} & 13 \\ \cline{2-4}
 & Cross-Site Request Forgery (CSRF) & \cite{carverandbosu2014,yuandfu2023,edmundsonetal2013} & 3 \\ \cline{2-4}
 & Command Injection & \cite{carverandbosu2014,yuandfu2023} & 2 \\ \hline

\multirow{5}{6cm}{\textbf{Improper Access Control (CWE-284)}}
 & Too Much Privilege & \cite{dibiaseatal2016} & 5 \\ \cline{2-4}
 & Weak Password-based Systems & \cite{dibiaseatal2016} & 1 \\ \cline{2-4}
 & Privilege / Sandbox Issues & \cite{dibiaseatal2016} & 2 \\ \cline{2-4}
 & Improper Authentication & \cite{alfadelandandnagy2023,wachiraphanandel2024,mahmoudetal2023} & 43 \\ \cline{2-4}
 & Improper Access Control & \cite{rajpauletal2021,carverandbosu2014,alfadelandandnagy2023,yuandfu2023} & 101 \\ \hline

\multirow{4}{6cm}{\textbf{Improper Adherence to Coding Standards (CWE-710)}}
 & Documentation Issues & \cite{alfadelandandnagy2023} & 14 \\ \cline{2-4}
 & Vulnerable Packages & \cite{alfadelandandnagy2023} & 3 \\ \cline{2-4}
 & Security Misconfiguration & \cite{carverandbosu2014} & 1 \\ \cline{2-4}
 & Potentially Dangerous Functions & \cite{rajpauletal2021} & 53 \\ \hline

\multirow{3}{6cm}{\textbf{Improper Check or Handling of Exceptional Conditions (CWE-703)}}
 & Catching All Exceptions & \cite{dibiaseatal2016} & 3 \\ \cline{2-4}
 & Crash & \cite{yuandfu2023,carverandbosu2014} & 142 \\ \cline{2-4}
 & Improper Exception Handling & \cite{rajpauletal2021} & 58 \\ \hline

\multirow{2}{6cm}{\textbf{Protection Mechanism Failure (CWE-693)}}
 & Failing to Protect Network Traffic & \cite{dibiaseatal2016} & 2 \\ \cline{2-4}
 & Encryption  & \cite{yuandfu2023,wachiraphanandel2024} & 24 \\ \hline

\multirow{2}{6cm}{\textbf{Incorrect Calculation (CWE-682)}}
 & Integer Overflows & \cite{dibiaseatal2016,carverandbosu2014,yuandfu2023,rajpauletal2021} & 86 \\ \cline{2-4}
 & Incorrect Calculation & \cite{rajpauletal2021} & 65 \\ \hline

\multirow{1}{6cm}{\textbf{C++ Catastrophes}}
 & C++-specific Vulnerabilities & \cite{dibiaseatal2016} & 8 \\ \hline

\multirow{1}{6cm}{\textbf{CWE-634 (Deprecated)}}
 & Weaknesses Affecting System Processes (IPC) & \cite{dibiaseatal2016} & 1 \\ \hline

\end{longtable}

\end{landscape}


\begin{landscape}
\small
\setlength{\tabcolsep}{4pt}
\begin{longtable}{|m{6.7cm}|p{5.8cm}|p{1.8cm}|p{1cm}|}
\caption{Missed or overlooked potential security vulnerabilities }
\label{tab:missedvuln}\\
\hline
\textbf{CWE Category (High level)} & \textbf{Vulnerability Name} & \textbf{Citations} & \textbf{Freq.} \\ \hline
\endfirsthead
\hline
\textbf{CWE Category (High level)} & \textbf{Vulnerability Name} & \textbf{Citations} & \textbf{Freq.} \\ \hline
\endhead
\endfoot
\hline
\endlastfoot

\multirow{17}{5.8cm}{\textbf{Improper Control of a Resource Through its Lifetime (CWE-664)}}
 & Buffer Overruns & \cite{dibiaseatal2016,Anetal2018} & 18 \\ \cline{2-4}
 & Information Leakage & \cite{dibiaseatal2016} & 1 \\ \cline{2-4}
 & Format String Problems & \cite{dibiaseatal2016} & 3 \\ \cline{2-4}
 & Magic URLs / Predictable Cookies & \cite{dibiaseatal2016} & 1 \\ \cline{2-4}
 & Incorrect Type Conversion or Cast & \cite{dibiaseatal2016,rajpauletal2021} & 18 \\ \cline{2-4}
 & Improper Control of a Resource in its Lifetime & \cite{dibiaseatal2016} & 2 \\ \cline{2-4}
 & Improper Synchronization & \cite{rajpauletal2021} & 13 \\ \cline{2-4}
 & Improper Resource Shutdown or Release & \cite{rajpauletal2021} & 4 \\ \cline{2-4}
 & Operation after Expiration or Release & \cite{rajpauletal2021} & 62 \\ \cline{2-4}
 & Improper Restriction within Memory Bounds & \cite{rajpauletal2021} & 16 \\ \cline{2-4}
 & Exposure of Resource to Wrong Sphere & \cite{rajpauletal2021} & 35 \\ \cline{2-4}
 & Session Expiration & \cite{fatemehandal2022} & 1 \\ \cline{2-4}
 & Memory Leak & \cite{fatemehandal2022,Anetal2018} & 3 \\ \cline{2-4}
 & Dangling Pointer & \cite{Anetal2018} & 1 \\ \cline{2-4}
 & Uninitialized Memory Read & \cite{Anetal2018} & 1 \\ \cline{2-4}
 & Incorrect Memory Allocation & \cite{Anetal2018} & 1 \\ \hline

\multirow{2}{5.8cm}{\textbf{Protection Mechanism Failure (CWE-693)}}
 & Failing to Protect Network Traffic & \cite{dibiaseatal2016} & 4 \\ \cline{2-4}
 & Insufficient Verification of Data Authenticity & \cite{rajpauletal2021} & 13 \\ \hline

\multirow{4}{5.8cm}{\textbf{Improper Access Control (CWE-284)}}
 & Too Much Privilege & \cite{dibiaseatal2016} & 5 \\ \cline{2-4}
 & Weak Password-based Systems & \cite{dibiaseatal2016} & 1 \\ \cline{2-4}
 & Improper Access Control & \cite{rajpauletal2021} & 58 \\ \cline{2-4}
 & Authentication & \cite{fatemehandal2022} & 1 \\ \hline

\multirow{5}{5.8cm}{\textbf{Improper Neutralization (CWE-707)}}
 & XSS (client side) & \cite{dibiaseatal2016} & 12 \\ \cline{2-4}
 & Command Injection & \cite{dibiaseatal2016} & 7 \\ \cline{2-4}
 & XSS (server side) & \cite{dibiaseatal2016} & 1 \\ \cline{2-4}
 & Improper Input Validation & \cite{rajpauletal2021} & 72 \\ \cline{2-4}
 & Data Processing Errors & \cite{rajpauletal2021} & 10 \\ \hline

\multirow{3}{5.8cm}{\textbf{Insufficient Control Flow Management (CWE-691)}}
 & Concurrency & \cite{Anetal2018} & 1 \\ \cline{2-4}
 & Incorrect Control Flow & \cite{Anetal2018} & 1 \\ \cline{2-4}
 & Race Conditions & \cite{dibiaseatal2016,fatemehandal2022,rajpauletal2021} & 15 \\ \hline

\multirow{1}{5.8cm}{\textbf{C++ Catastrophes}}
 & C++-specific Vulnerabilities & \cite{dibiaseatal2016} & 33 \\ \hline

\multirow{3}{5.8cm}{\textbf{Improper Adherence to Coding Standards (CWE-710)}}
 & Null Pointer Dereference & \cite{Anetal2018} & 1 \\ \cline{2-4}
 & Double Free & \cite{Anetal2018} & 1 \\ \cline{2-4}
 & Potentially Dangerous Function & \cite{dibiaseatal2016} & 2 \\ \hline

\multirow{1}{9cm}{\parbox[t]{9cm}{\RaggedRight\textbf{Improper Check or Handling of \\ Exceptional Conditions (CWE-703)}}}
 & Improper Check or Handling of Exceptional Conditions & \cite{rajpauletal2021,Anetal2018} & 22 \\ \hline

\multirow{2}{5.8cm}{\textbf{Incorrect Calculation (CWE-682)}}
 & Incorrect Calculation & \cite{rajpauletal2021,Anetal2018} & 17 \\ \cline{2-4}
 & Integer Overflows & \cite{dibiaseatal2016,Anetal2018} & 3 \\ \hline

\multirow{1}{5.8cm}{\textbf{CWE-634 (Deprecated)}}
 & Weaknesses Affecting System Processes (IPC) & \cite{dibiaseatal2016} & 1 \\ \hline

\end{longtable}

\end{landscape}


\begin{center}
\small
\begin{longtable}{|p{0.7cm}|p{5cm}|p{5cm}|p{1cm}|}
\caption{Primary studies included in this SLR.}
\label{tab:primary_studies} \\

\hline
\textbf{Sl. No.} & \textbf{Title} & \textbf{Authors} & \textbf{Year} \\
\hline
\endfirsthead

\hline
\textbf{Sl. No.} & \textbf{Title} & \textbf{Authors} & \textbf{Year} \\
\hline
\endhead

S1 & An Empirical Study of Static Analysis Tools for Secure Code Review~\cite{charoenwet2024} & Wachiraphan Charoenwet, Patanamon Thongtanunam, Van-Thuan Pham, Christoph Treude & 2024 \\ \hline

S2 & Toward effective secure code reviews: an empirical study of security-related coding weaknesses~\cite{wachiraphanandel2024} & Wachiraphan Charoenwet, Patanamon Thongtanunam, Van-Thuan Pham, Christoph Treude & 2024 \\ \hline

S3 & Empirical analysis of security vulnerabilities in Python packages~\cite{mahmoudetal2023} & Mahmoud Alfadel, Diego Elias Costa, Emad Shihab & 2023 \\ \hline

S4 & Security Defect Detection via Code Review: A Study of the OpenStack and Qt Communities~\cite{yuandfu2023} & Jiaxin Yu, Liming Fu, Peng Liang, Amjed Tahir, Mojtaba Shahin & 2023 \\ \hline

S5 & Empirical analysis of security-related code reviews in npm packages~\cite{alfadelandandnagy2023} & Mahmoud Alfadel, Nicholas Alexandre Nagy, Diego Elias Costa, Rabe Abdalkareem, Emad Shihab & 2023 \\ \hline

S6 & Less is More: Supporting Developers in Vulnerability Detection during Code Review~\cite{brazandchristian2022} & Larissa Braz, Christian Aeberhard, Gül Çalikli, Alberto Bacchelli & 2022 \\ \hline

S7 & Software Security during Modern Code Review: The Developer’s Perspective~\cite{brazandbacchelli2022} & Larissa Braz, Alberto Bacchelli & 2022 \\ \hline

S8 & Which bugs are missed in code reviews: An empirical study on SmartSHARK dataset~\cite{fatemehandal2022} & Fatemeh Khoshnoud, Ali Rezaei Nasab, Zahra Toudeji, Ashkan Sami & 2022 \\ \hline

S9 & Why Security Defects Go Unnoticed during Code Reviews? A Case-Control Study of the Chromium OS Project~\cite{rajpauletal2021} & Rajshakhar Paul, Asif Kamal Turzo, Amiangshu Bosu & 2021 \\ \hline

S10 & A systematic literature review and taxonomy of modern code review~\cite{Davilaetal2021} & Nicole Davila, Ingrid Nunes & 2021 \\ \hline

S11 & An Empirical Investigation of Relevant Changes and Automation Needs in Modern Code Review~\cite{panichellaetal2020} & Sebastiano Panichella, Nik Zaugg & 2020 \\ \hline

S12 & Why Did This Reviewed Code Crash? An Empirical Study of Mozilla Firefox~\cite{Anetal2018} & Le An, Foutse Khomh, Shane McIntosh, Marco Castelluccio & 2018 \\ \hline

S13 & A Large-Scale Study of Modern Code Review and Security in Open Source Projects~\cite{thompsonetal2017} & Christopher Thompson, David Wagner & 2017 \\ \hline

S14 & A security perspective on code review: The case of Chromium~\cite{dibiaseatal2016} & Marco di Biase, Magiel Bruntink, Alberto Bacchelli & 2016 \\ \hline

S15 & Identifying the characteristics of vulnerable code changes: an empirical study~\cite{carverandbosu2014} & Amiangshu Bosu, Jeffrey C. Carver, Munawar Hafiz, Patrick Hilley, Derek Janni & 2014 \\ \hline

S16 & An Empirical Investigation of Socio-technical Code Review Metrics and Security Vulnerabilities~\cite{meneelyatel2014} & Andrew Meneely, Alberto C. Rodriguez Tejeda, Brian Spates, Shannon Trudeau, DanielleNeuberger, Katherine Whitlock, Christopher Ketant, Kayla Davis & 2014 \\ \hline

S17 & The Impact of Code Review Coverage and Code Review Participation on Software Quality: a case study of the qt, VTK, and ITK projects~\cite{mcintosh2014} & Shane McIntosh, Yasutaka Kamei, Bram Adams, Ahmed E. Hassan & 2014 \\ \hline

S18 & Expectations, outcomes, and challenges of modern code review~\cite{bacchbird2013} & Alberto Bacchelli, Christian Bird & 2013 \\ \hline

S19 & An Empirical Study on the Effectiveness of Security Code Review~\cite{edmundsonetal2013} & Anne Edmundson, Brian Holtkamp, Emanuel Rivera, Matthew Finifter, Adrian Mettler, David Wagner & 2013 \\ \hline

S20 & Static security analysis based on input-related software faults~\cite{nagyandal2009} & Csaba Nagy, Spiros Mancoridis & 2009 \\ \hline



\end{longtable}
\end{center}

%% file: cas-refs.bib
@inproceedings{Ackerman1984,
  title={Software inspections and the industrial production of software},
  author={A. F. Ackerman and Priscilla J. Fowler and Robert G. Ebenau},
  year={1984},
  pages={13-40},
  booktitle = {Elsevier North-Holland, Inc.},
  url={https://api.semanticscholar.org/CorpusID:59830250}
}

@ARTICLE{ackerman1989,
  author={Ackerman, A.F. and Buchwald, L.S. and Lewski, F.H.},
  journal={IEEE Software}, 
  title={Software inspections: an effective verification process}, 
  year={1989},
  volume={6},
  number={3},
  pages={31-36},
  doi={10.1109/52.28121}}

@INPROCEEDINGS{bacchbird2013,
  author={Bacchelli, Alberto and Bird, Christian},
  booktitle={2013 35th International Conference on Software Engineering (ICSE)}, 
  title={Expectations, outcomes, and challenges of modern code review}, 
  year={2013},
  volume={},
  number={},
  pages={712-721},
  doi={10.1109/ICSE.2013.6606617}}

@book{mcgraw2006,
author = {McGraw, Gary},
title = {Software Security: Building Security In},
year = {2006},
isbn = {0321356705},
publisher = {Addison-Wesley Professional},
}

@inproceedings{carverandbosu2014,
author = {Bosu, Amiangshu and Carver, Jeffrey C. and Hafiz, Munawar and Hilley, Patrick and Janni, Derek},
title = {Identifying the characteristics of vulnerable code changes: an empirical study},
year = {2014},
isbn = {9781450330565},
publisher = {Association for Computing Machinery},
address = {New York, NY, USA},
url = {https://doi.org/10.1145/2635868.2635880},
doi = {10.1145/2635868.2635880},
booktitle = {Proceedings of the 22nd ACM SIGSOFT International Symposium on Foundations of Software Engineering},
pages = {257–268},
numpages = {12},
location = {Hong Kong, China},
series = {FSE 2014}
}

@ARTICLE{fagan1976,
  author={Fagan, M. E.},
  journal={IBM Systems Journal}, 
  title={Design and code inspections to reduce errors in program development}, 
  year={1976},
  volume={15},
  number={3},
  pages={182-211},
  doi={10.1147/sj.153.0182}}

@article{mantylaandcasper2009,
author = {Mantyla, Mika V. and Lassenius, Casper},
title = {What Types of Defects Are Really Discovered in Code Reviews?},
year = {2009},
issue_date = {May 2009},
publisher = {IEEE Press},
volume = {35},
number = {3},
issn = {0098-5589},
url = {https://doi.org/10.1109/TSE.2008.71},
doi = {10.1109/TSE.2008.71},
journal = {IEEE Trans. Softw. Eng.},
month = {may},
pages = {430–448},
numpages = {19},
}

@inproceedings{bellerandbacchelli2014,
author = {Beller, Moritz and Bacchelli, Alberto and Zaidman, Andy and Juergens, Elmar},
title = {Modern code reviews in open-source projects: which problems do they fix?},
year = {2014},
isbn = {9781450328630},
publisher = {Association for Computing Machinery},
address = {New York, NY, USA},
url = {https://doi.org/10.1145/2597073.2597082},
doi = {10.1145/2597073.2597082},
booktitle = {Proceedings of the 11th Working Conference on Mining Software Repositories},
pages = {202–211},
numpages = {10},
location = {Hyderabad, India},
series = {MSR 2014}
}

@inproceedings{peterandbird2013,
author = {Rigby, Peter C. and Bird, Christian},
title = {Convergent contemporary software peer review practices},
year = {2013},
isbn = {9781450322379},
publisher = {Association for Computing Machinery},
address = {New York, NY, USA},
url = {https://doi.org/10.1145/2491411.2491444},
doi = {10.1145/2491411.2491444},
booktitle = {Proceedings of the 2013 9th Joint Meeting on Foundations of Software Engineering},
pages = {202–212},
numpages = {11},
location = {Saint Petersburg, Russia},
series = {ESEC/FSE 2013}
}

@article{cohen2010modern,
  title={Modern code review},
  author={Cohen, Jason},
  journal={Making Software: What Really Works, and Why We Believe It},
  pages={329--336},
  year={2010}
}

@inproceedings{mcintosh2014,
author = {McIntosh, Shane and Kamei, Yasutaka and Adams, Bram and Hassan, Ahmed E.},
title = {The impact of code review coverage and code review participation on software quality: a case study of the qt, VTK, and ITK projects},
year = {2014},
isbn = {9781450328630},
publisher = {Association for Computing Machinery},
address = {New York, NY, USA},
url = {https://doi.org/10.1145/2597073.2597076},
doi = {10.1145/2597073.2597076},
booktitle = {Proceedings of the 11th Working Conference on Mining Software Repositories},
pages = {192–201},
numpages = {10},
location = {Hyderabad, India},
series = {MSR 2014}
}

@InProceedings{edmundsonetal2013,
author="Edmundson, Anne
and Holtkamp, Brian
and Rivera, Emanuel
and Finifter, Matthew
and Mettler, Adrian
and Wagner, David",
editor="J{\"u}rjens, Jan
and Livshits, Benjamin
and Scandariato, Riccardo",
title="An Empirical Study on the Effectiveness of Security Code Review",
booktitle="Engineering Secure Software and Systems",
year="2013",
publisher="Springer Berlin Heidelberg",
address="Berlin, Heidelberg",
pages="197--212",
isbn="978-3-642-36563-8"
}

@misc{kitchenham2007slrguidelines,
  title={Guidelines for performing systematic literature reviews in software engineering},
  author={Kitchenham, Barbara and Charters, Stuart and others},
  year={2007},
  publisher={UK}
}

@INPROCEEDINGS{dibiaseatal2016,
  author={di Biase, Marco and Bruntink, Magiel and Bacchelli, Alberto},
  booktitle={2016 IEEE 16th International Working Conference on Source Code Analysis and Manipulation (SCAM)}, 
  title={A Security Perspective on Code Review: The Case of Chromium}, 
  year={2016},
  volume={},
  number={},
  pages={21-30},
  doi={10.1109/SCAM.2016.30}}

@inproceedings{rajpauletal2021,
author = {Paul, Rajshakhar and Turzo, Asif Kamal and Bosu, Amiangshu},
title = {Why Security Defects Go Unnoticed during Code Reviews? A Case-Control Study of the Chromium OS Project},
year = {2021},
isbn = {9781450390859},
publisher = {IEEE Press},
url = {https://doi.org/10.1109/ICSE43902.2021.00124},
doi = {10.1109/ICSE43902.2021.00124},
booktitle = {Proceedings of the 43rd International Conference on Software Engineering},
pages = {1373–1385},
numpages = {13},
location = {Madrid, Spain},
series = {ICSE '21}
}

@inproceedings{fatemehandal2022,
author = {Khoshnoud, Fatemeh and Nasab, Ali Rezaei and Toudeji, Zahra and Sami, Ashkan},
title = {Which bugs are missed in code reviews: an empirical study on SmartSHARK dataset},
year = {2022},
isbn = {9781450393034},
publisher = {Association for Computing Machinery},
address = {New York, NY, USA},
url = {https://doi.org/10.1145/3524842.3527997},
doi = {10.1145/3524842.3527997},
booktitle = {Proceedings of the 19th International Conference on Mining Software Repositories},
pages = {137–141},
numpages = {5},
location = {Pittsburgh, Pennsylvania},
series = {MSR '22}
}

@inproceedings{nagyandal2009,
author = {Nagy, Csaba and Mancoridis, Spiros},
title = {Static Security Analysis Based on Input-Related Software Faults},
year = {2009},
isbn = {9780769535890},
publisher = {IEEE Computer Society},
address = {USA},
url = {https://doi.org/10.1109/CSMR.2009.51},
doi = {10.1109/CSMR.2009.51},
booktitle = {Proceedings of the 2009 European Conference on Software Maintenance and Reengineering},
pages = {37–46},
numpages = {10},
series = {CSMR '09}
}

@article{Davilaetal2021,
  title={A Systematic Literature Review and Taxonomy of Modern Code Review},
  author={Nicole Davila and Ingrid Nunes},
  journal={J. Syst. Softw.},
  year={2021},
  volume={177},
  pages={110951},
  url={https://api.semanticscholar.org/CorpusID:232240592}
}

@INPROCEEDINGS{Anetal2018,
  author={An, Le and Khomh, Foutse and Mcintosh, Shane and Castelluccio, Marco},
  booktitle={2018 25th Asia-Pacific Software Engineering Conference (APSEC)}, 
  title={Why Did This Reviewed Code Crash? An Empirical Study of Mozilla Firefox}, 
  year={2018},
  volume={},
  number={},
  pages={396-405},
  doi={10.1109/APSEC.2018.00054}}

@inproceedings{brazandbacchelli2022,
author = {Braz, Larissa and Bacchelli, Alberto},
title = {Software security during modern code review: the developer’s perspective},
year = {2022},
isbn = {9781450394130},
publisher = {Association for Computing Machinery},
address = {New York, NY, USA},
url = {https://doi.org/10.1145/3540250.3549135},
doi = {10.1145/3540250.3549135},
booktitle = {Proceedings of the 30th ACM Joint European Software Engineering Conference and Symposium on the Foundations of Software Engineering},
pages = {810–821},
numpages = {12},
location = {Singapore, Singapore},
series = {ESEC/FSE 2022}
}

@article{meneelyatel2014,
author = {Meneely, Andrew and Tejeda, Alberto and Spates, Brian and Trudeau, Scott and Neuberger, Danielle and Whitlock, Katherine and Ketant, Christopher and Davis, Kayla},
year = {2014},
month = {11},
pages = {37-44},
title = {An empirical investigation of socio-technical code review metrics and security vulnerabilities},
journal = {6th International Workshop on Social Software Engineering, SSE 2014 - Proceedings},
doi = {10.1145/2661685.2661687}
}

@inproceedings{thompsonetal2017,
author = {Thompson, Christopher and Wagner, David},
title = {A Large-Scale Study of Modern Code Review and Security in Open Source Projects},
year = {2017},
isbn = {9781450353052},
publisher = {Association for Computing Machinery},
address = {New York, NY, USA},
url = {https://doi.org/10.1145/3127005.3127014},
doi = {10.1145/3127005.3127014},
booktitle = {Proceedings of the 13th International Conference on Predictive Models and Data Analytics in Software Engineering},
pages = {83–92},
numpages = {10},
location = {Toronto, Canada},
series = {PROMISE}
}

@INPROCEEDINGS{kononenkoetal2015,
  author={Kononenko, Oleksii and Baysal, Olga and Guerrouj, Latifa and Cao, Yaxin and Godfrey, Michael W.},
  booktitle={2015 IEEE International Conference on Software Maintenance and Evolution (ICSME)}, 
  title={Investigating code review quality: Do people and participation matter?}, 
  year={2015},
  volume={},
  number={},
  pages={111-120},
  doi={10.1109/ICSM.2015.7332457}}

@article{panichellaetal2020,
author = {Panichella, Sebastiano and Zaugg, Nik},
title = {An Empirical Investigation of Relevant Changes and Automation Needs in Modern Code Review},
year = {2020},
issue_date = {Nov 2020},
publisher = {Kluwer Academic Publishers},
address = {USA},
volume = {25},
number = {6},
issn = {1382-3256},
url = {https://doi.org/10.1007/s10664-020-09870-3},
doi = {10.1007/s10664-020-09870-3},
journal = {Empirical Softw. Engg.},
month = {nov},
pages = {4833–4872},
numpages = {40}
}

@article{howard2006,
author = {Howard, M.A.},
year = {2006},
month = {08},
pages = {74 - 79},
title = {A process for performing security code reviews},
volume = {4},
journal = {Security and Privacy, IEEE},
doi = {10.1109/MSP.2006.84}
}

@inproceedings{brazandchristian2022,
author = {Braz, Larissa and Aeberhard, Christian and \c{C}alikli, G\"{u}l and Bacchelli, Alberto},
title = {Less is more: supporting developers in vulnerability detection during code review},
year = {2022},
isbn = {9781450392211},
publisher = {Association for Computing Machinery},
address = {New York, NY, USA},
url = {https://doi.org/10.1145/3510003.3511560},
doi = {10.1145/3510003.3511560},
booktitle = {Proceedings of the 44th International Conference on Software Engineering},
pages = {1317–1329},
numpages = {13},
location = {Pittsburgh, Pennsylvania},
series = {ICSE '22}
}

@misc{OWASPAbout2013,
  author = {{OWASP} {F}oundation},
  title = {About {OWASP}},
  howpublished = {online},
  note = {[Accesssed online 04-11-2013]},
  url = "https://www.owasp.org/index.php/About\_OWASP",
  year = 2013
}

@ARTICLE{bosuandcarver2017,
  author={Bosu, Amiangshu and Carver, Jeffrey C. and Bird, Christian and Orbeck, Jonathan and Chockley, Christopher},
  journal={IEEE Transactions on Software Engineering}, 
  title={Process Aspects and Social Dynamics of Contemporary Code Review: Insights from Open Source Development and Industrial Practice at Microsoft}, 
  year={2017},
  volume={43},
  number={1},
  pages={56-75},
  doi={10.1109/TSE.2016.2576451}}

@misc{NIST_report,
  key = 	 {CVE-2008-1368},
  title = 	 {National Institute of Standards and Technology, National Vulnerability Database},
  institution  = {NIST},
  month =     {March},
  year = 	 2024,
  url = {http://https://nvd.nist.gov/vuln/search/statistics?form\_type=Basic\&results\_type=statistics\&search\_type=all\&isCpeNameSearch=false\#}
}

@book{chessandjacob2007,
author = {Chess, Brian and West, Jacob},
title = {Secure programming with static analysis},
year = {2007},
isbn = {9780321424778},
publisher = {Addison-Wesley Professional},
edition = {First},

}

@inproceedings{Thongtanunametal2015,
author = {Thongtanunam, Patanamon and Mcintosh, Shane and Hassan, Ahmed E. and Iida, Hajimu},
year = {2015},
month = {03},
pages = {},
title = {Investigating Code Review Practices in Defective Files: An Empirical Study of the Qt System},
doi = {10.1109/MSR.2015.23}
}

@article{shinetal2011,
author = {Shin, Yonghee and Meneely, Andrew and Williams, Laurie and Osborne, Jason},
year = {2011},
month = {11},
pages = {772-787},
title = {Evaluating Complexity, Code Churn, and Developer Activity Metrics as Indicators of Software Vulnerabilities},
volume = {37},
journal = {IEEE Trans. Software Eng.},
doi = {10.1109/TSE.2010.81}
}

@misc{Equifax,
  author = {Suraj Srinivasan and Quinn Pitcher and Jonah S. Goldberg},
  title = {Data Breach at Equifax},
  year = {2017},
  publisher = {Harvard Business School},
  note = {Harvard Business School Case 118-031, Revised April 2019},
  url = {https://www.hbs.edu/faculty/Pages/item.aspx?num=53509}
}

@ARTICLE{Heartbleed_2014,
  author={Wheeler, David A.},
  journal={Computer}, 
  title={Preventing Heartbleed}, 
  year={2014},
  volume={47},
  number={8},
  pages={80-83},
  doi={10.1109/MC.2014.217}}

@misc{charoenwet2024,
      title={An Empirical Study of Static Analysis Tools for Secure Code Review}, 
      author={Wachiraphan Charoenwet and Patanamon Thongtanunam and Van-Thuan Pham and Christoph Treude},
      year={2024},
      eprint={2407.12241},
      archivePrefix={arXiv},
      primaryClass={cs.SE},
      url={https://arxiv.org/abs/2407.12241}, 
}

@article{MacGrwa_2008,
author = {McGraw, Gary},
title = {Automated Code Review Tools for Security},
year = {2008},
issue_date = {December 2008},
publisher = {IEEE Computer Society Press},
address = {Washington, DC, USA},
volume = {41},
number = {12},
issn = {0018-9162},
journal = {Computer},
month = dec,
pages = {108–111},
numpages = {4},
}

@inproceedings{shahedandhassan2011,
author = {Zaman, Shahed and Adams, Bram and Hassan, Ahmed E.},
title = {Security versus performance bugs: a case study on Firefox},
year = {2011},
isbn = {9781450305747},
publisher = {Association for Computing Machinery},
address = {New York, NY, USA},
url = {https://doi.org/10.1145/1985441.1985457},
doi = {10.1145/1985441.1985457},
booktitle = {Proceedings of the 8th Working Conference on Mining Software Repositories},
pages = {93–102},
numpages = {10},
location = {Waikiki, Honolulu, HI, USA},
series = {MSR '11}
}

@article{Aurumandwohlin2002,
author = {Aurum, Aybüke and Petersson, Håkan and Wohlin, Claes},
year = {2002},
month = {09},
pages = {133 - 154},
title = {State-of-the-art: Software inspections after 25 years},
volume = {12},
journal = {Software Testing, Verification and Reliability},
doi = {10.1002/stvr.243}
}

@INPROCEEDINGS{camiloandmeneely2015,
  author={Camilo, Felivel and Meneely, Andrew and Nagappan, Meiyappan},
  booktitle={2015 IEEE/ACM 12th Working Conference on Mining Software Repositories}, 
  title={Do Bugs Foreshadow Vulnerabilities? A Study of the Chromium Project}, 
  year={2015},
  volume={},
  number={},
  pages={269-279},
  doi={10.1109/MSR.2015.32}}

@INPROCEEDINGS{meneelyandsrinivasan2013,
  author={Meneely, Andrew and Srinivasan, Harshavardhan and Musa, Ayemi and Tejeda, Alberto Rodríguez and Mokary, Matthew and Spates, Brian},
  booktitle={2013 ACM / IEEE International Symposium on Empirical Software Engineering and Measurement}, 
  title={When a Patch Goes Bad: Exploring the Properties of Vulnerability-Contributing Commits}, 
  year={2013},
  volume={},
  number={},
  pages={65-74},
 doi={10.1109/ESEM.2013.19}}

@inproceedings{ashisandjagannath2013,
author = {Ashish, A. and Aghav, Jagannath},
year = {2013},
month = {07},
pages = {1-7},
title = {Automated techniques and tools for program analysis: Survey},
isbn = {978-1-4799-3926-8},
booktitle = {2013 4th International Conference on Computing, Communications and Networking Technologies, ICCCNT 2013},
doi = {10.1109/ICCCNT.2013.6726693}
}

@inproceedings{johnsonandrobert2013,
author = {Johnson, Brittany and Song, Yoonki and Murphy-Hill, Emerson and Bowdidge, Robert},
title = {Why don't software developers use static analysis tools to find bugs?},
year = {2013},
isbn = {9781467330763},
publisher = {IEEE Press},
booktitle = {Proceedings of the 2013 International Conference on Software Engineering},
pages = {672–681},
numpages = {10},
location = {San Francisco, CA, USA},
series = {ICSE '13}
}

@article{arthoandarmin2005,
author = {Artho, Cyrille and Biere, Armin},
year = {2005},
month = {05},
pages = {3-14},
title = {Combined Static and Dynamic Analysis},
volume = {131},
journal = {Electr. Notes Theor. Comput. Sci.},
doi = {10.1016/j.entcs.2005.01.018}
}

@article{boehmetal2001,
author = {Boehm, Barry and Basili, Victor R.},
title = {Software Defect Reduction Top 10 List},
year = {2001},
issue_date = {January 2001},
publisher = {IEEE Computer Society Press},
address = {Washington, DC, USA},
volume = {34},
number = {1},
issn = {0018-9162},
url = {https://doi.org/10.1109/2.962984},
doi = {10.1109/2.962984},
journal = {Computer},
month = jan,
pages = {135–137},
numpages = {3}
}

@InProceedings{Munaiahandmeyers2017,
author="Munaiah, Nuthan and Meyers, Benjamin S. and Alm, Cecilia O. and Meneely, Andrew and Murukannaiah, Pradeep K. and Prud'hommeaux, Emily
and Wolff, Josephine and Yu, Yang",
editor="Bodden, Eric
and Payer, Mathias
and Athanasopoulos, Elias",
title="Natural Language Insights from Code Reviews that Missed a Vulnerability",
booktitle="Engineering Secure Software and Systems",
year="2017",
publisher="Springer International Publishing",
address="Cham",
pages="70--86",
isbn="978-3-319-62105-0"
}

@article{alfadelandandnagy2023,
author = {Alfadel, Mahmoud and Nagy, Nicholas Alexandre and Costa, Diego Elias and Abdalkareem, Rabe and Shihab, Emad},
title = {Empirical analysis of security-related code reviews in npm packages},
year = {2023},
issue_date = {Sep 2023},
publisher = {Elsevier Science Inc.},
address = {USA},
volume = {203},
number = {C},
issn = {0164-1212},
url = {https://doi.org/10.1016/j.jss.2023.111752},
doi = {10.1016/j.jss.2023.111752},
journal = {J. Syst. Softw.},
month = sep,
numpages = {19}
}

@misc{davisjam2021vulnregex,
  author = {Davis, Jam},
  title = {vulnregex-detector: Detect vulnerable regexes in your project},
  year = {2021},
  url = {https://github.com/davisjam/vuln-regex-detector},
  note = {Accessed: 2024-11-14}
}

@article{wachiraphanandel2024,
author = {Charoenwet, Wachiraphan and Thongtanunam, Patanamon and Pham, Van-Thuan and Treude, Christoph},
title = {Toward effective secure code reviews: an empirical study of security-related coding weaknesses},
year = {2024},
issue_date = {Jul 2024},
publisher = {Kluwer Academic Publishers},
address = {USA},
volume = {29},
number = {4},
issn = {1382-3256},
url = {https://doi.org/10.1007/s10664-024-10496-y},
doi = {10.1007/s10664-024-10496-y},
journal = {Empirical Softw. Engg.},
month = jun,
numpages = {47}
}

@INPROCEEDINGS {yuandfu2023,
author = { Yu, Jiaxin and Fu, Liming and Liang, Peng and Tahir, Amjed and Shahin, Mojtaba },
booktitle = { 2023 ACM/IEEE International Symposium on Empirical Software Engineering and Measurement (ESEM) },
title = {{ Security Defect Detection via Code Review: A Study of the OpenStack and Qt Communities }},
year = {2023},
volume = {},
ISSN = {},
pages = {1-12},
doi = {10.1109/ESEM56168.2023.10304852},
url = {https://doi.ieeecomputersociety.org/10.1109/ESEM56168.2023.10304852},
publisher = {IEEE Computer Society},
address = {Los Alamitos, CA, USA},
month =Oct}

@ARTICLE{McGraw2004,
  author={McGraw, G.},
  journal={IEEE Security \& Privacy}, 
  title={Software security}, 
  year={2004},
  volume={2},
  number={2},
  pages={80-83},
  doi={10.1109/MSECP.2004.1281254}}

@misc{mitre_cwe,
  author       = {{MITRE Corporation}},
  title        = {{Common Weakness Enumeration}},
  year         = {2024},
  url          = {https://cwe.mitre.org/},
  note         = {[Online; accessed 13-April-2025]}
}

@article{Telangetal2007,
author = {Telang, Rahul and Wattal, Sunil},
title = {An Empirical Analysis of the Impact of Software Vulnerability Announcements on Firm Stock Price},
year = {2007},
issue_date = {August 2007},
publisher = {IEEE Press},
volume = {33},
number = {8},
issn = {0098-5589},
url = {https://doi.org/10.1109/TSE.2007.70712},
doi = {10.1109/TSE.2007.70712},
journal = {IEEE Trans. Softw. Eng.},
month = aug,
pages = {544–557},
numpages = {14}
}

@article{mahmoudetal2023,
author = {Alfadel, Mahmoud and Costa, Diego Elias and Shihab, Emad},
title = {Empirical analysis of security vulnerabilities in Python packages},
year = {2023},
issue_date = {May 2023},
publisher = {Kluwer Academic Publishers},
address = {USA},
volume = {28},
number = {3},
issn = {1382-3256},
url = {https://doi.org/10.1007/s10664-022-10278-4},
doi = {10.1007/s10664-022-10278-4},
journal = {Empirical Softw. Engg.},
month = mar,
numpages = {37}
}

@inproceedings{brownetal2020,
author = {Brown, Tom B. and Mann, Benjamin and Ryder, Nick and Subbiah, Melanie and Kaplan, Jared and Dhariwal, Prafulla and Neelakantan, Arvind and Shyam, Pranav and Sastry, Girish and Askell, Amanda and Agarwal, Sandhini and Herbert-Voss, Ariel and Krueger, Gretchen and Henighan, Tom and Child, Rewon and Ramesh, Aditya and Ziegler, Daniel M. and Wu, Jeffrey and Winter, Clemens and Hesse, Christopher and Chen, Mark and Sigler, Eric and Litwin, Mateusz and Gray, Scott and Chess, Benjamin and Clark, Jack and Berner, Christopher and McCandlish, Sam and Radford, Alec and Sutskever, Ilya and Amodei, Dario},
title = {Language models are few-shot learners},
year = {2020},
isbn = {9781713829546},
publisher = {Curran Associates Inc.},
address = {Red Hook, NY, USA},
booktitle = {Proceedings of the 34th International Conference on Neural Information Processing Systems},
articleno = {159},
numpages = {25},
location = {Vancouver, BC, Canada},
series = {NIPS '20}
}

@inproceedings{longetal2022,
author = {Ouyang, Long and Wu, Jeff and Jiang, Xu and Almeida, Diogo and Wainwright, Carroll L. and Mishkin, Pamela and Zhang, Chong and Agarwal, Sandhini and Slama, Katarina and Ray, Alex and Schulman, John and Hilton, Jacob and Kelton, Fraser and Miller, Luke and Simens, Maddie and Askell, Amanda and Welinder, Peter and Christiano, Paul and Leike, Jan and Lowe, Ryan},
title = {Training language models to follow instructions with human feedback},
year = {2022},
isbn = {9781713871088},
publisher = {Curran Associates Inc.},
address = {Red Hook, NY, USA},
booktitle = {Proceedings of the 36th International Conference on Neural Information Processing Systems},
articleno = {2011},
numpages = {15},
location = {New Orleans, LA, USA},
series = {NIPS '22}
}

@inproceedings{xiaetal2023,
author = {Xia, Chunqiu Steven and Wei, Yuxiang and Zhang, Lingming},
title = {Automated Program Repair in the Era of Large Pre-Trained Language Models},
year = {2023},
isbn = {9781665457019},
publisher = {IEEE Press},
url = {https://doi.org/10.1109/ICSE48619.2023.00129},
doi = {10.1109/ICSE48619.2023.00129},
booktitle = {Proceedings of the 45th International Conference on Software Engineering},
pages = {1482–1494},
numpages = {13},
location = {Melbourne, Victoria, Australia},
series = {ICSE '23}
}

@article{lietal2024,
author = {Li, Haonan and Hao, Yu and Zhai, Yizhuo and Qian, Zhiyun},
year = {2024},
month = {04},
pages = {474-499},
title = {Enhancing Static Analysis for Practical Bug Detection: An LLM-Integrated Approach},
volume = {8},
journal = {Proceedings of the ACM on Programming Languages},
doi = {10.1145/3649828}
}

@article{steenhoeketal2024,
  publtype={informal},
  author={Benjamin Steenhoek and Md Mahbubur Rahman and Monoshi Kumar Roy and Mirza Sanjida Alam and Earl T. Barr and Wei Le},
  title={A Comprehensive Study of the Capabilities of Large Language Models for Vulnerability Detection},
  year={2024},
  cdate={1704067200000},
  journal={CoRR},
  volume={abs/2403.17218},
  url={https://doi.org/10.48550/arXiv.2403.17218}
}

@misc{khareetal2024,
      title={Understanding the Effectiveness of Large Language Models in Detecting Security Vulnerabilities}, 
      author={Avishree Khare and Saikat Dutta and Ziyang Li and Alaia Solko-Breslin and Rajeev Alur and Mayur Naik},
      year={2024},
      eprint={2311.16169},
      archivePrefix={arXiv},
      primaryClass={cs.CR},
      url={https://arxiv.org/abs/2311.16169}, 
}

@misc{xuetal20252025,
      title={Hallucination is Inevitable: An Innate Limitation of Large Language Models}, 
      author={Ziwei Xu and Sanjay Jain and Mohan Kankanhalli},
      year={2025},
      eprint={2401.11817},
      archivePrefix={arXiv},
      primaryClass={cs.CL},
      url={https://arxiv.org/abs/2401.11817}, 
}

@article{Kuhrmann2016,
         title={On the pragmatic design of literature studies in software engineering: an experience-based guideline},
         author={M. Kuhrmann and Daniel Méndez Fernández and M. Daneva},
         journal={Empirical Software Engineering},
         year={2016},
         volume={22},
         pages={2852 - 2891},
         doi={10.1007/s10664-016-9492-y}
         }

@article{Palazzi2019,
    title={Online division of labour: emergent structures in Open Source Software},
    author={M. Palazzi and Jordi Cabot and J. L. Cánovas Izquierdo and Albert Solé-Ribalta and Javier Borge-Holthoefer},
    journal={Scientific Reports},
    year={2019},volume={9},
    doi={10.1038/s41598-019-50463-y}}

@article{Guizani2021,
    title={The Long Road Ahead: Ongoing Challenges in Contributing to Large OSS Organizations and What to Do},
    author={Mariam Guizani and Amreeta Chatterjee and Bianca Trinkenreich and Mary Evelyn May and Geraldine J. Noa-Guevara and Liam James Russell and Griselda G. Cuevas Zambrano and Daniel Izquierdo-Cortazar and I. Steinmacher and M. Gerosa and A. Sarma},
    journal={Proceedings of the ACM on Human-Computer Interaction},
    year={2021},
    volume={5},
    pages={1 - 30},
    doi={10.1145/3479551}
    }

@article{Xu2025,
    title={The Structure of Cross-National Collaboration in Open-Source Software Development},
    author={H. Xu and Katy Yu and Hao He and Hongbo Fang and Bogdan Vasilescu and Patrick S. Park},
    journal={Proceedings of the 34th ACM International Conference on Information and Knowledge Management},
    year={2025},
    doi={10.1145/3746252.3761237}
    }

@article{Trinkenreich2020,
    title={Hidden Figures: Roles and Pathways of Successful OSS Contributors},
    author={Bianca Trinkenreich and Mariam Guizani and I. Wiese and A. Sarma and I. Steinmacher},
    journal={Proceedings of the ACM on Human-Computer Interaction},year={2020},
    volume={4},
    pages={1 - 22},
    doi={10.1145/3415251}
    }

@article{Osborne2024,
    title={Characterising Open Source Co-opetition in Company-hosted Open Source Software Projects: The Cases of PyTorch, TensorFlow, and Transformers},author={Cailean Osborne and Farbod Daneshyan and Runzhi He and Hengzhi Ye and Yuxia Zhang and Minghui Zhou},
    journal={Proceedings of the ACM on Human-Computer Interaction},
    year={2024},
    volume={9},
    pages={1 - 30},
    doi={10.1145/3710944}
    }

@article{Linaker2025,
    title={Public sector open source software projects - How is development organized?},
    author={Johan Linåker and Björn Lundell and Francisco Servant and Jonas Gamalielsson and Sachiko Muto and Gregorio Robles},
    journal={Empirical Software Engineering},
    year={2025},
    volume={30},
    doi={10.1007/s10664-025-10626-0}
    }

@misc{mitre_cwe_1000,
  title        = {{CWE}-1000: Research Concepts (Version 4.20)},
  author       = {{MITRE Corporation}},
  howpublished = {\url{https://cwe.mitre.org/data/definitions/1000.html}},
  note         = {Accessed: May 2026}
}
